\PassOptionsToPackage{capitalize,nameinlink}{cleveref}

\PassOptionsToPackage{table}{xcolor}
\PassOptionsToPackage{svgnames}{xcolor}

\PassOptionsToPackage{final}{microtype}

\documentclass[sigconf, pdfa]{acmart}

\setcopyright{acmlicensed}
\copyrightyear{2018}
\acmYear{2018}
\acmDOI{XXXXXXX.XXXXXXX}
\acmConference[Conference acronym 'XX]{Make sure to enter the correct
  conference title from your rights confirmation email}{June 03--05,
  2018}{Woodstock, NY}
\acmISBN{978-1-4503-XXXX-X/2018/06}

\newcommand{\paperTitle}{This is Going to Sound Crazy, But What If We Used Large Language Models
to \underline{Boost} Automatic Database Tuning Algorithms By Leveraging Prior History? We Will Find 
Better Configurations More Quickly Than Retraining From Scratch!}

\newcommand{\paperShortTitle}{Booster}

\newcommand{\pg}{PostgreSQL\xspace}

\newcommand{\captionTitle}[1]{\textbf{#1} --}

\usepackage[a-2b]{pdfx}

\definecolor{RowGray}{gray}{0.95}

\usepackage{algorithm,algpseudocode}
\usepackage{enumitem}
\usepackage{graphicx}
\usepackage{balance}
\usepackage{stfloats}
\usepackage{xspace}
\usepackage{tikz}
\usepackage{tabularx}
\usepackage{enumitem}
\usepackage{lipsum}
\usepackage{subcaption}
\usepackage[export]{adjustbox}
\usepackage{cleveref}
\usepackage[framemethod=tikz]{mdframed}
\usepackage{soul}
\usepackage{fancyvrb}

\makeatletter
\renewcommand\paragraph{\@startsection{paragraph}{4}{\parindent}%
  {0.03in}%
  {-3.5\p@}%
  {\ACM@NRadjust{\@parfont\@adddotafter}}}
\makeatother

\Crefname{section}{Sec.}{Secs.}
\Crefname{table}{Tab.}{Tabs.}

\newcommand{\dbPGTune}{PGTune\xspace}
\newcommand{\dbDTA}{DTA\xspace}
\newcommand{\dbLambda}{LambdaTune\xspace}

\newcommand{\lambdagpt}{{GPT-4o}\xspace}
\newcommand{\dbUniTune}{UniTune\xspace}
\newcommand{\dbPDA}{\textsf{P+DTA+AS}\xspace}
\newcommand{\dbPX}{\textsf{Proto-X}\xspace}
\newcommand{\dbLT}{\textsf{$\lambda$-Tune+AS}\xspace}

\newcommand{\techniqueembed}{\textsf{Voyage 3 Large}\xspace}
\newcommand{\techniqueprompt}{\textsf{Llama 3.1-8B-Instruct Q4-K-M}\xspace}
\newcommand{\techniquepromptbone}{\textsf{Llama 3.1-8B-Instruct}\xspace}

\newcommand{\qwen}{{Qwen3-1.7B}\xspace}
\newcommand{\othreemini}{{o3-mini}\xspace}
\newcommand{\textlarge}{{text-3-large}\xspace}
\newcommand{\nomiccode}{{Nomic Code}\xspace}

\newcommand{\dbIndex}[1]{\textsf{I#1}\xspace}
\newcommand{\dbQuery}[1]{\textsf{Q#1}\xspace}
\newcommand{\dbConfig}[1]{\textsf{C#1}\xspace}
\newcommand{\dbQCItem}[2]{\textsf{Q#1}-\textsf{C#2}\xspace}
\newcommand{\dbIV}[1]{\textsf{IV#1}\xspace}
\newcommand{\dbAPI}[1]{\texttt{#1}\xspace}

\newcommand{\dbTechnique}{\textsf{Booster}\xspace}

\newcommand{\dbQC}{\textsf{QConfig}\xspace}
\newcommand{\dbSanitize}{\textsf{Sanitizer}\xspace}

\newcommand{\dbAS}{AutoSteer\xspace}

\newcommand{\dbPGHintPlan}{pg\_hint\_plan\xspace}
\newcommand{\dbHypoPG}{HypoPG\xspace}
\newcommand{\dbHypoExec}{HypoExec\xspace}

\newcommand{\CMU}{Carnegie Mellon University\xspace}

\definecolor{red}{rgb}{1,0,0}
\definecolor{color-will}{rgb}{0,0.5,0.5}

\definecolor{todo-color}{rgb}{1,0,0}

\definecolor{color-wan}{rgb}{0.8,0,0}

\newcommand{\furl}[1]{\footnote{\url{#1}}}

\newcommand*\circled[1]{%
    \tikz[baseline=(char.base)]{\node[shape=circle,fill=gray,draw,inner sep=0.5pt] (char) {\color{white}\textsf{#1}};}%
}%

\begin{document}

\title[\paperShortTitle]{\paperTitle}

\newcommand{\superscript}[1]{\ensuremath{^{#1}}}
\def\xCMU{}

\author{William Zhang}
\orcid{0000-0002-9392-6683}
\email{wz2@cs.cmu.edu}
\affiliation{
    \institution{\CMU}
}

\author{Wan Shen Lim}
\orcid{0000-0003-1508-2080}
\email{wanshenl@cs.cmu.edu}
\affiliation{
    \institution{\CMU}
}

\author{Andrew Pavlo}
\orcid{0000-0001-6040-6991}
\email{pavlo@cs.cmu.edu}
\affiliation{
    \institution{\CMU}
}


\begin{abstract}

Tuning database management systems (DBMSs) is challenging due to trillions of possible
configurations and evolving workloads. Recent advances in tuning have led to breakthroughs in
optimizing over the possible configurations. However, due to their design and inability to leverage
query-level historical insights, existing automated tuners struggle to adapt and re-optimize the
DBMS when the environment changes (e.g., workload drift, schema transfer).

This paper presents the \dbTechnique framework that assists existing tuners in adapting
to environment changes (e.g., drift, cross-schema transfer). \dbTechnique structures historical
artifacts into query-configuration contexts, prompts large language models (LLMs) to suggest
configurations for each query based on relevant contexts, and then composes the query-level
suggestions into a holistic configuration with beam search. With multiple OLAP workloads, we
evaluate \dbTechnique's ability to assist different state-of-the-art tuners (e.g.,
cost-/machine learning-/LLM-based) in adapting to environment changes. By composing recommendations
derived from query-level insights, \dbTechnique assists tuners in discovering configurations that
are up to 74\% better and in up to 4.7$\times$ less time than the alternative
approach of
continuing to tune from historical configurations.

\end{abstract}


\maketitle

%


\section{Introduction}
\label{sec:introduction}

Configuring a database management system (DBMS) for modern data-intensive applications has become
increasingly difficult for two reasons. First, DBMSs expose trillions of options~\cite{lim23}
(e.g., knobs, indexes, query hints). Second, these applications are constantly evolving, with
changes in data access patterns, query types, and load intensities, amongst
others~\cite{renen24redshift,pavlo21}. Automated tuners ensure that the DBMS remains in its optimal
configuration as the environment (e.g., workload, data contents) changes over time~\cite{pavlo17}.

Optimizing a DBMS for a fixed workload has been well explored by the academic community.
Early research focused on heuristics and cost-based
search techniques~\cite{pgtune,storm06,dexter,surajit98autoadmin,surajit20anytime}.
The last decade saw the rise of using machine learning (ML) to tune single aspects of
the DBMS (e.g., knobs~\cite{vanaken17,zhang19cdb,kanellis22,huang2025e2e},
indexes~\cite{kossmann20,wu22budget,siddiqui22}) or across all aspects of the
DBMS~\cite{zhang24holon}. Recent advances have focused on large language model (LLM)-based
methods~\cite{giannakouris25,huang2025e2e}.

Existing research has shown the efficacy of these tuners in finding beneficial configurations for a
fixed workload. However, these tuners are unable to adapt to environment changes (e.g., workload
drift, different schemas) due to their inherent designs. For example, heuristic
tuners~\cite{pgtune,dexter} cannot incorporate historical knowledge due to their fixed algorithms.
ML-tuners~\cite{kossmann22,zhang23unitune,zhang24holon} struggle as environment changes (e.g.,
query \dbQuery{1} no longer exists) corrupt their internal representations (e.g., how they represent
\dbQuery{1}'s tunables). These design issues prevent tuners from effectively exploiting query-level
semantics (e.g., query plan) to learn from historical knowledge. Consequently, these tuners take
longer to re-optimize after environment changes and end up stuck in subpar configurations.

Given this, we introduce the \textbf{\dbTechnique} framework that
assists an existing tuner (e.g., cost-/ML-/LLM-based) in adapting to environment changes.
\dbTechnique first organizes historical tuning artifacts into structured insights. When the DBMS
environment changes, \dbTechnique obtains configurations derived from relevant
insights based on each individual query's semantics, composes them into a holistic
configuration~\cite{zhang24holon} (i.e., its findings), and injects those findings into the
tuner for further refinement. We evaluate \dbTechnique's ability to assist state-of-the-art
tuners in adapting to new DBMS environments for OLAP workloads.
Our experiments with \pg show that
\dbTechnique assists tuners in finding configurations that improve DBMS performance by up to
74\% in up to 4.7$\times$ less time
compared to the alternative approach of continuing to tune from history.

We lay out the paper as follows. We provide background into existing tuners' limitations in
adapting to environment changes and ML techniques underpinning our method in \cref{sec:background}.
We then provide an overview of \dbTechnique in \cref{sec:overview}, followed by
describing its execution phases in \cref{sec:experience,sec:composition}.
We then discuss how to integrate the framework with existing tuners in
\cref{sec:integration}. We evaluate \dbTechnique's ability to assist tuners in adapting in
\cref{sec:eval,sec:eval-sensitivity}.

\section{Background}
\label{sec:background}

An autonomous self-driving DBMS~\cite{pavlo17,zhou21dbmind,kossmann20selfdriving} optimizes itself
without human intervention.
Guided by the user's objectives (e.g., minimize runtime), the DBMS deploys \textit{tuners} to
optimize itself for a given workload.
These tuners iteratively explore and experiment with different configurations to find the optimal
settings for the system. These include cost-based search and ML-based
tuners that target individual DBMS aspects (e.g.,
knobs~\cite{pgtune,vanaken17,zhang19cdb,kanellis22},
physical design~\cite{kossmann20,wu22budget,siddiqui22,ding19aiai,ahmed20}),
holistic tuners that reason across multiple DBMS aspects~\cite{zhang24holon}, and large language
model (LLM)-based tuners~\cite{giannakouris25,huang2025e2e,yao2025qoptllm}.
We first define adaptivity for tuners and then discuss existing tuners along with their challenges
in adapting to different deployments.

\subsection{Adaptivity for Tuners}
\label{sec:adaptivity}

As a tuner optimizes DBMSs, 
it acquires experience that includes the tuner's reasoning, explored configurations, and observed
behavior of the DBMS.
\textit{Adaptivity} reflects a tuner's ability to reuse prior experience when
tuning new scenarios, broadly categorized as \textit{transfer} and \textit{drift}. We will discuss
each separately.

\paragraph{Transfer Scenario} This addresses the case where a tuner transfers experiences from past
DBMS deployments to a previously unseen deployment. This ranges from cases where the historical
and target deployments are the same to cases where the schemas differ. For example, after tuning
a TPC-H~\cite{tpch} instance, the tuner transfers its experiences to optimize a
TPC-DS~\cite{tpcds} instance.

\paragraph{Drift Scenario} This covers scenarios where a tuner optimizes a DBMS deployment
whose characteristics change over time~\cite{li22warper,renen24redshift,wang21}. These
include changes to query parameters, query templates, 
query volume (i.e., load spikes), hardware (e.g., instance upgrades), and underlying data (e.g.,
\verb|BULK|~\verb|INSERT|).
These drifts are common in practice, with Redshift
observing 50\% of their production clusters have 50\% of queries repeating exactly (i.e., same
query template and parameters)
daily~\cite{renen24redshift}.

\subsection{Existing Tuning Frameworks}
\label{sec:existing-tuners}


We next provide an overview of existing state-of-the-art tuners, which we group into three
categories: (1) heuristic and cost-based, (2) ML-based, and (3) LLM-based.

\paragraph{Heuristic and Cost-based Tuners}
These tuners execute a fixed algorithm to explore
configurations~\cite{pgtune,dexter,surajit20anytime,anneser23}
obtained with heuristics. They then rely on query plan costs via a ``what-if''
mechanism~\cite{surajit98autoadmin} to guide the search process.

\paragraph{ML-based and Holistic Tuners}
These tuners rely on ML techniques to tune the DBMS. While some of these tuners may only target
specific aspects (e.g., knobs, indexes) of the
DBMS~\cite{vanaken17,zhang19cdb,siddiqui22,kossmann22},
holistic tuners~\cite{zhang24holon} reason across all aspects.
These tuners rely on an internal model to find beneficial configurations through trial and
error. These models capture deployment aspects (e.g., workload, schema) as bits in the
models' representation. For instance, a specific bit output by the model may instruct the
tuner to build an index~\cite{wang21udo,kossmann22}, set a system knob~\cite{vanaken17,zhang19cdb},
or apply a query hint~\cite{zhang24holon}.

\paragraph{LLM-based Tuners}
Recent large language model (LLM)-based tuners rely on off-the-shelf models~\cite{giannakouris25}
(e.g., \lambdagpt~\cite{gpt4o}) 
or models fine-tuned on prior experience~\cite{huang2025e2e}.
They then instruct the LLM with \textit{prompts}.
For instance, a prompt can instruct the LLM to assume the role of a database administrator and
output a JSON configuration that optimizes the workload.
These tuners sample multiple configurations from the LLM and select the best one.

\subsection{Tuner Adaptivity Challenges}
\label{sec:limited-adapting}
Despite the prevalence of transfer and drift scenarios in practice, existing tuners
are unable to adequately adapt to environment changes due to their designs.
As we now discuss, their design limitations are due to two core issues:

\paragraph{Rigidity and Brittleness}
This issue is present in cost-based and ML-based tuners. For both, 
identifying relevant experiences is critical to discovering beneficial
configurations~\cite{vanaken17,zhang22hpo,zhang23opadvisor,li24r2}.
Even if relevant experience is accurately identified, problems remain.

Cost-based tuners lack a mechanism to apply prior insights 
to guide the search. For instance, if prior experience reveals an adverse interaction between
two indexes, the tuner should prioritize exploring configurations that exclude those two indexes.

ML-based tuners must re-optimize whenever their internal models' representations
change due to the environment (e.g., new query, new indexable column).
To compensate for this, researchers proposed using relevant experience to pre-train
the internal model~\cite{li19}. However, pre-training remains infeasible due to potential
differences with the target environment 
and the overhead of re-evaluating historical experiences (i.e., configurations) to obtain
performance values for the target schema and workload.

\paragraph{Workload Granularity}
Existing tuners adapt on a workload granularity. Rather than tune from scratch, tuners use
workload-level telemetry (e.g., DBMS metrics) to identify and start from some historical
configuration (i.e., workload mapping~\cite{vanaken17}). However, this mapping prevents combining
individual query insights across configurations. For instance, configurations \dbConfig{1} and
\dbConfig{2} may optimize different workload queries more effectively.
Without a query-level adaptation method based on query semantics
(e.g., plan)~\cite{li19,bianchi24db2une}, tuners cannot
compose the best aspects of \dbConfig{1} and \dbConfig{2} together.

LLM-based tuners ignore opportunities to augment the prompt with targeted
experience to guide the model's suggestion process on a query-by-query basis. For instance, if a
query has been tuned before, the tuner could augment the prompt with past attempts (e.g., query
hints, indexes) to provide the LLM with further context to generate more
effective configurations~\cite{lewis20rag,li24r2,chen25debugrag}.

\vspace*{0.05in}

\begin{figure}[t!]
    \begin{subfigure}{\linewidth}
        \centering
        \setlength{\fboxsep}{0pt}
        \framebox[.95\linewidth]{\includegraphics[height=0.12in]{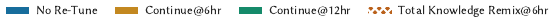}}

        \includegraphics[width=\linewidth]{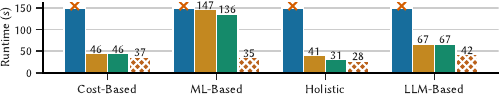}
        \captionsetup{aboveskip=0pt}
        \caption{Transfer Scenario from (TPC-H $\rightarrow$ DSB)}
        \label{fig:motivation-transfer}
    \end{subfigure}

    \vspace{0.5em}
    \begin{subfigure}{\linewidth}
        \centering

        \includegraphics[width=\linewidth]{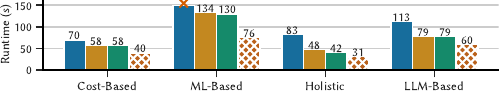}
        \captionsetup{aboveskip=0pt}
        \caption{Drift Scenario (50\% of Parameters Changed for DSB Queries)}
        \label{fig:motivation-drift}
    \end{subfigure}

    \caption{
        \captionTitle{Tuner Adaptivity Challenges}
        \pg runtimes achieved 
        by tuners in transfer 
        and drift 
        scenarios for three configurations:
        (1) \textit{no re-tune},
        (2) \textit{continue} tuning from the best historical configuration for 6hrs and
        12hrs, and
        (3) \textit{remix} prior knowledge and then tune for 6hrs.
        Each tuner has access
        to historical artifacts: TPC-H (transfer) and prior DSB (drift).
    }
    \label{fig:motivation}
\end{figure}

To illustrate how these problems hinder DBMS tuners, we ran workloads on \pg with
two scenarios:
(1) an experience transfer from TPC-H~\cite{tpch} to DSB~\cite{dsb} and
(2) a previously tuned DSB workload undergoing a parameter drift (i.e., 50\% of
queries have
different parameters).
We deploy a (1) cost-based tuner~\cite{pgtune,surajit20anytime,anneser23},
(2) ML-based tuner~\cite{zhang23unitune},
(3) holistic tuner~\cite{zhang24holon},
and (4) LLM-based tuner~\cite{giannakouris25,anneser23}. Each tuner has access to their
historical
artifacts: TPC-H for transfer scenario and the previous DSB for drift scenario.

As shown in \cref{fig:motivation-transfer} and \cref{fig:motivation-drift}, both scenarios benefit
from continuing to tune from the best historical configuration compared to \textbf{No Re-Tune}.
However, \textbf{Continue} falls short of what is achievable by \textbf{Total Knowledge Remix}.
By fully remixing historical tuning knowledge, tuners discover configurations that are
\textbf{13--74\%} better than those found by \textbf{Continue}.
To achieve this for a range of tuners, we
propose leveraging recent advances in LLMs to reason about individual
queries and adapt them to different environments.

\subsection{LLM-based Query Adaptation}
\label{sec:llm-adaptation}

Recent research in LLMs has shown their capabilities in
optimization~\cite{tao2025,yao2025qoptllm,li24r2,giannakouris25}
and
schema understanding~\cite{zhang24react,yan24gidcl,xiu24qartisan}.
However, providing curated information to the LLM remains
unsolved~\cite{gao2024ragsurvey,zhang2024instruct,minaee2025llm}. We provide
an exposition into incorporating tuning knowledge into LLMs through four techniques:
(1) workload-level prompts, (2) workload-level fine-tune, (3) query fine-tune, and
(4) combining enriched query-level prompts with a composition mechanism.

\paragraph{Workload-Level Prompts} This technique (\textbf{WL-Prompt}) invokes an
off-the-shelf LLM with a workload-level prompt that describes the tuning task, the workload, and
additional information (e.g., DBMS telemetry, workload summary) as context~\cite{lewis20rag}.
This technique utilizes the LLM's innate abilities and pre-trained knowledge to generate
configurations rather than historical knowledge.

\paragraph{Workload-Level Fine-Tune} This technique (\textbf{WL-FT}) fine-tunes a LLM
by using historical experiences to alter the LLM's weights~\cite{huang2025e2e}. Similarly to
\textbf{WL-Prompt}, this technique then invokes the fine-tuned LLM to obtain candidate
configurations. By fine-tuning, this technique improves the LLM's ability to generate
more effective configurations for the target workload.

\paragraph{Query Fine-Tune} In comparison to \textbf{WL-FT}, this technique (\textbf{Q-FT})
fine-tunes a LLM on queries. Although this enables the LLM to relate queries to
configurations, it requires prompting the LLM with a query prompt. As such, rather
than generating configurations for the entire workload, the LLM generates configurations for
each query that are then combined into a holistic configuration~\cite{zhang24holon}.

\paragraph{Enriched Query-Level Prompts and Composition} This technique (\textbf{Enrich Q+Compose})
is agnostic to whether the LLM is fine-tuned or not. On a per-query basis, this technique
enriches the query prompt with prior tuning attempts based on the query (e.g., SQL text, plan) and
then instructs the LLM to generate configurations with those historical
references~\cite{zhang2024instruct,lewis20rag,chen25debugrag}.
The technique then uses a composition mechanism to combine the query-level configurations into a
holistic configuration while resolving conflicts (e.g., different knob values, conflicting
indexes). In doing so, this technique achieves \textbf{Total Knowledge Remix} by reasoning over
all available historical knowledge (e.g., past configurations, Internet).

\begin{figure}[t!]
    \includegraphics[width=\linewidth]{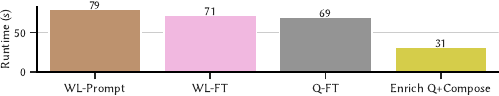}
    \caption{
        \captionTitle{LLM-based Query Adaptation}
        DSB workload runtime achieved by
        prompting an off-the-shelf LLM (\textbf{WL-Prompt}), fine-tuning a LLM with historical
        knowledge at workload (\textbf{WL-FT}) and query (\textbf{Q-FT}) granularities, and
        \textbf{Enrich Q+Compose} that combines query prompts enriched with historical attempts
        and a composition mechanism. All techniques utilize experience generated over 12hrs by a
        holistic tuner tuning a DSB workload where 50\% of queries have different parameters.
    }
    \label{fig:motivation-ft}
\end{figure}

\vspace*{0.05in}

To understand their efficacy, we first obtain historical experience by using a
holistic tuner to optimize a DSB workload for 12hrs. We then evaluate the previous techniques
on using this 
experience to adapt to a drifted workload where 50\% of the queries have different parameters.
As shown in \cref{fig:motivation-ft}, \textbf{WL-Prompt} performs the worst as it
does not use prior experience. \textbf{WL-FT} and \textbf{Q-FT} have limited improvement
over \textbf{WL-Prompt} due to issues around granularity and combining query
configurations, respectively.
In contrast, \textbf{Enrich Q+Compose} finds drastically better configurations by
exploiting relevant knowledge on a per-query basis and resolving conflicts with a composition
mechanism. \textbf{Enrich Q+Compose} forms the basis of our method to assist existing tuners in
adapting to environment changes by remixing historical knowledge with LLMs.

\section{Overview}
\label{sec:overview}

\begin{figure}[t!]
    \includegraphics[width=0.95\columnwidth]{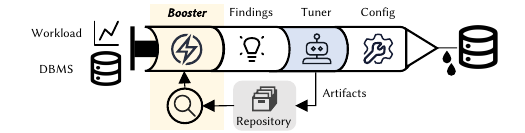}
    \caption{
        \captionTitle{\dbTechnique Overview}
        The framework integrates with an existing tuner to improve its adaptivity to environment
        changes. 
        \dbTechnique analyzes the artifact repository and injects its findings (e.g., start
        configuration) into the tuner. The ``injected'' tuner then refines the configuration
        and stores its artifacts into the repository for further analysis.
    }
    \label{fig:overview}
\end{figure}

\begin{figure}[t!]
    \includegraphics[width=0.95\columnwidth]{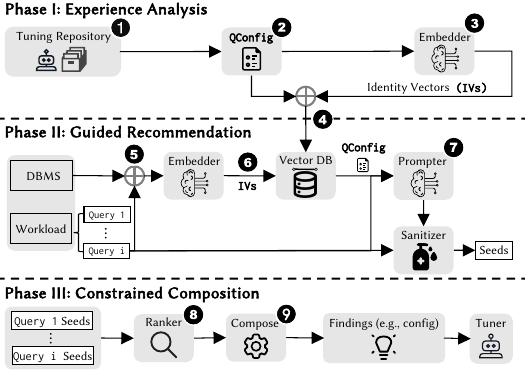}
    \caption{
        \captionTitle{\dbTechnique Architecture} An overview of the framework's three phases.
        In Phase I, \dbTechnique analyzes historical tuning artifacts. In Phase II,
        \dbTechnique generates 
        candidate configurations (i.e., \textit{seeds}) for each
        query with a LLM. In Phase III, \dbTechnique then composes each query's seeds
        into a holistic configuration that it then provides to the tuner being assisted.
    }
    \label{fig:arch}
\end{figure}

We present how the \dbTechnique framework integrates with an existing tuner
(see \cref{sec:existing-tuners}) to improve its ability to adapt to environment changes in
\cref{fig:overview}.
Based on the workload and DBMS, \dbTechnique analyzes the repository of artifacts and injects
its findings (e.g., start configuration) into the tuner. The injected tuner then further refines
the configuration and 
updates the repository with new artifacts.

We next describe how \dbTechnique utilizes this repository to derive a
holistic configuration~\cite{zhang24holon} (i.e., its findings) for injecting into the tuner.
As shown in \cref{fig:arch}, \dbTechnique's operation is divided into three phases. In the first
phase, it 
monitors the repository for new artifacts generated by the tuner
and analyzes them for insights (\cref{sec:overview-phase1}). When the DBMS environment changes,
\dbTechnique uses these insights to optimize the DBMS. In the second phase, \dbTechnique receives
the user's target workload and the target DBMS (\cref{sec:overview-phase2}). For each query in the
workload, \dbTechnique identifies relevant experiences and utilizes them to generate candidate
configurations
with a LLM. Then in the third phase, \dbTechnique combines these query-level candidate
configurations with a constrained composition mechanism (\cref{sec:overview-phase3}).
Lastly, \dbTechnique provides its findings
to the tuner for further refinement. We next discuss each phase in more detail.

\subsection{Phase I: Experience Analysis}
\label{sec:overview-phase1}

In this phase, \circled{1} \dbTechnique 
monitors a repository of tuning experiences.
This repository accumulates artifacts generated by a tuner as it optimizes a deployment.
Although these artifacts contain substantial information about the DBMS (e.g., workloads,
configurations' efficacy), they are not readily searchable and differ across tuners.
To standardize these artifacts into a searchable format, \dbTechnique first \circled{2} parses them
into 
query-configuration (\dbQC) objects that represent how a query behaves
under a specific configuration. Each \dbQC includes but is not limited to the relevant schema, the
query's execution plan, the configuration (e.g., knobs, indexes), and links to other related
\dbQC{s} for structural organization. For instance, \dbTechnique can link \dbQC{s}
together in chronological order if the \dbQC{s} refer to the same query.

With these \dbQC{s}, \dbTechnique next makes them searchable by leveraging recent advances in
LLMs' ability to understand tabular data, schemas, and
SQL queries~\cite{zhang24react,yan24gidcl,li24r2,xiu24qartisan}. For each \dbQC, \circled{3}
\dbTechnique utilizes an embedder (e.g., \techniqueembed~\cite{voyageai}) to generate fixed-length
identity vectors (i.e., embeddings~\cite{rege23embeddings}) based on text documents derived from
the \dbQC. For instance, it can derive them by combining the schema and query plan
or by combining the anonymized schema and SQL text.
\dbTechnique combines these
identity vectors with the \dbQC and \circled{4} loads them into a vector database for querying
in Phase II. We elaborate further in \cref{sec:experience}.

\subsection{Phase II: Guided Recommendation}
\label{sec:overview-phase2}

At the start of this phase, \dbTechnique is now optimizing the DBMS.
\dbTechnique receives the target workload and a connection to an offline
environment for safe exploration~\cite{ma20active,lim23}.
For each query in the workload, \circled{5} \dbTechnique obtains relevant information from the DBMS
(e.g., schema, plan) and uses an embedder to generate identity vectors in the
same manner as Phase I. With each query's identity vectors, \dbTechnique \circled{6}
retrieves relevant historical \dbQC{s} from the vector database, combines them
with the target query into the prompt, and \circled{7} invokes the LLM to
recommend configurations based on the \dbQC{s}~\cite{lewis20rag,chen25debugrag}.
As the LLM may suggest invalid configurations (e.g., illegal knob value, invalid index),
\dbTechnique passes these suggested configurations and the relevant \dbQC{s} through the
\dbSanitize to obtain 
configurations for each query that can be deployed. As these
configurations target a single query, we refer to them as \textit{query seeds}.
We elaborate further in \cref{sec:experience}.

\subsection{Phase III: Constrained Composition}
\label{sec:overview-phase3}

After obtaining the seeds for each query,
\dbTechnique then combines them into a
holistic configuration~\cite{zhang24holon}. However, seed configurations may specify different
parameters (e.g., number of parallel workers, indexes) 
that conflict when 
combined. To mitigate this, \dbTechnique adopts a two-step process to generate
holistic configurations from multiple query seeds. \circled{8} \dbTechnique first 
ranks all the query seeds and then \circled{9} runs a beam search algorithm, a variant
of best-first search~\cite{chen23loger,pearl84,azizi23elpis},
to identify performant holistic configurations. This algorithm terminates when
a terminal condition (e.g., elapsed time) is reached.
We elaborate on this further in \cref{sec:composition}.
Finally,
\dbTechnique injects its discoveries into the tuner being assisted.

\section{Experience Analysis \& Recommendations}
\label{sec:experience}

\dbTechnique uses \dbQC{s} to guide how its LLM recommends configurations. We now
discuss (1) how the framework constructs \dbQC{s}, (2) how it augments the prompt with relevant
\dbQC{s}, and (3) how \dbSanitize produces seed configurations for each query.

\subsection{\dbQC Construction}
\label{sec:qc-construction}

\begin{figure}[t!]
    \includegraphics[width=0.95\columnwidth]{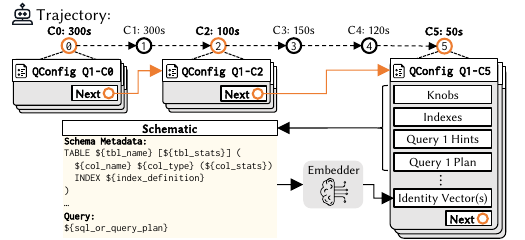}
    \caption{
        \captionTitle{\dbQC Construction} \dbTechnique generates \dbQC objects from interesting
        configurations (e.g., configurations that improved the user's objective function):
        \dbConfig{0}, \dbConfig{2}, and \dbConfig{5}. As \dbQCItem{1}{5} illustrates, each \dbQC
        contains information (e.g., knobs, plan) about a specific query in a configuration, a link
        to a downstream configuration (e.g., \dbQCItem{1}{2} to \dbQCItem{1}{5}), and multiple
        identity vectors (i.e., embeddings) obtained by passing different schematics through
        an embedder.
    }
    \label{fig:qc-construct}
\end{figure}

During Phase I, \dbTechnique mines the repository of artifacts generated by the tuner.
\dbTechnique analyzes the configurations explored over time (i.e., \textit{trajectory}) to identify
interesting configurations (e.g., ones that improve the user's objective function).
On a per-query basis, \dbTechnique
constructs a \dbQC from each interesting configuration. As shown in \cref{fig:qc-construct},
this \dbQC contains both the DBMS's configuration (e.g., knobs, indexes) and
query-specific information (e.g., SQL text, plan). The \dbQC also contains references
to downstream related (i.e., same query) \dbQC{s}. By default, \dbTechnique links each \dbQC
to the successive \dbQC that it constructs from the next configuration the tuner discovers
chronologically.
In this example, 
\dbQCItem{1}{0} links to \dbQCItem{1}{2}, which then links to \dbQCItem{1}{5}.

\dbTechnique cannot use these \dbQC{s} for vector-based similarity
search~\cite{kuffo25pdx,fan24}. Instead, \dbTechnique must first derive a fixed-length vector
representation
from each \dbQC before it can use distance functions (e.g., cosine distance, euclidean
distance) for similarity search. Recent research has proposed using embedders
(e.g., \techniqueembed~\cite{voyageai})
for natural language to SQL~\cite{fan24} and root cause analysis~\cite{ouyang25rca}.
These embedders map input texts to fixed-length vectors where
related texts are located nearby~\cite{chen25debugrag}.
\dbTechnique obtains a \dbQC's identity vectors by constructing representative texts
from the \dbQC and passing those texts through an embedder.

As shown in \cref{fig:qc-construct}, \dbTechnique builds multiple \textit{schematics} from each
\dbQC. These schematics capture a query's semantics, ranging from high-level information
about what data the query accesses to low-level details of how the DBMS will execute it.
Each schematic comprises two parts: (1) schema metadata and (2) the query. The schema metadata
describes the referenced relations and columns, utilized indexes, and additional statistics
(e.g., number of relation tuples, number of distinct values for each column). The query
component describes what the DBMS executes, which can take the form of the SQL query, the query
template, or query plan. \dbTechnique generates schematics based on all three, along with
variants derived by anonymizing table and column names.
It then passes each schematic through an embedder to obtain an identity vector (i.e.,
embedding) and loads the assembled \dbQC into a vector database. 

\subsection{\dbQC Guided Recommendation}
\label{sec:guided-recommendation}

\begin{figure}[t!]
    \includegraphics[width=0.95\columnwidth]{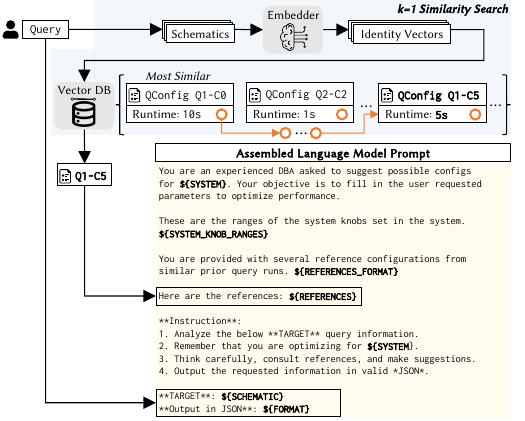}
    \caption{
        \captionTitle{Prompt Augmentation with k=1 Relevant \dbQC} Based on the query,
        \dbTechnique derives identity vectors in a similar process to
        \cref{fig:qc-construct}.
        It then retrieves a ranked list of \dbQC{s} based on similarity (i.e., Euclidean
        distance), takes the most similar \dbQC, and follows its links to obtain
        the most performant downstream \dbQC.
        \dbTechnique then enriches the prompt with the downstream \dbQC (\dbQC{Q1-C5}) and
        prompts the LLM for suggested configurations.
    }
    \label{fig:gr-pipeline}
\end{figure}

Using the vector database from Phase I, \dbTechnique in Phase II
obtains candidate configurations with a LLM.
For a given user query, \dbTechnique generates
schematics and computes identity vectors (\dbIV{s}) using the same process and embedder as in Phase
I.
With each query's \dbIV, \dbTechnique retrieves similar \dbQC from the vector store by minimizing
the euclidean
distance to each \dbQC's identity vector of the same schematic type. For example, assume
\dbTechnique uses the anonymized query plan to build schematic \verb|S| and obtains the query's
\dbIV{1} from it. When searching the vector store, \dbTechnique only considers the identity vector
from each \dbQC that is also obtained with \verb|S|. This ensures that \dbTechnique does not
erroneously rank \dbQC{s} based on 
differences arising from schematic type.

However, the most semantically similar \dbQC may not be the best reference for the LLM.
Consider the case where the DBMS is in the stock configuration and \dbTechnique is
analyzing a previously tuned query. In this case, the most semantically similar \dbQC{$^\prime$}
is derived from the same query in the same stock configuration.
Yet, this \dbQC{$^\prime$} lacks any
guidance (e.g., query knobs to set, indexes to build) for a LLM to exploit. Instead,
a more relevant \dbQC{$^\star$} with guidance exists downstream
(i.e., discovered later in time) of
\dbQC{$^\prime$}. Based on this observation, we make a core change to the retrieval process:
from each retrieved \dbQC, \dbTechnique follows its links to fetch the most performant downstream
\dbQC.
It then provides these downstream \dbQC{s} as references to the LLM.

Due to LLM context window limitations (i.e., prompt and response
length)~\cite{liu24lost,jin2024longcontext}, \dbTechnique truncates the references to the
top-$k$ \dbQC,
packages the reference \dbQC{s} into the prompt, and instructs the LLM to reason through those
references (Instruction \#3 in \cref{fig:gr-pipeline})~\cite{kojima22,lewis20rag} to suggest
configurations. Based on empirical trials, \dbTechnique sets $k$ to a default value of 2.

\subsection{Recommendation Sanitization}
\label{sec:recommendation-sanitization}

Prior techniques~\cite{huang2025e2e,giannakouris25} use the configurations suggested by the LLM
``as-is''.
However, these suggestions may be incomplete, contain invalid indexes and parameters (e.g., a knob
from a different DBMS), or request to match parameter values to the references.
To correct these errors,
\dbTechnique's \dbSanitize cleans the suggested configurations using each query's \dbQC references.

\dbTechnique forces the LLM to output valid JSON
to avoid the
complexity of extracting
relevant configuration
snippets.  \dbSanitize first constructs a preliminary list from three sources: (1) the LLM's
suggestions, (2) the \dbQC references, and (3) additional configurations obtained by filling in any
missing parameters in the LLM's suggestions with values from \dbQC references.

As this preliminary list may occupy a locally suboptimal region, \dbSanitize introduces a
limited degree of exploration. 
From each preliminary
configuration, \dbSanitize derives additional candidates in two ways to produce
diverse query plans:
(1) augment indexes based on static analysis of the query's predicates and
(2) permuting the query knobs (e.g., turn off sorting). 
For example, in \pg, turning off sorting while keeping other query knobs (e.g., access method
hints) the same tends to produce different yet also performant plans.
We provide a further sensitivity analysis in \cref{sec:sensitivity-dk}.

\dbSanitize then removes invalid selections (e.g., unknown columns, invalid hints) from each
candidate configuration before obtaining its minimal form. 
For example,
\dbSanitize uses a ``what-if'' mechanism~\cite{surajit98autoadmin} to identify indexes used by
and to remove unused indexes from each candidate. Another example is query hints that allow
multiple possibilities. For instance, \pg's \verb|NoSeqScan(t)|~\cite{pg_hint} hint requests the
DBMS to use an index scan if possible.
\dbSanitize modifies these hints based on the query plan to their more specific form (e.g, 
\verb|IndexScan(t)|). Finally, it preserves all
configurations that result in unique query plans as the query's seeds.

\section{Constrained Composition}
\label{sec:composition}

Once \dbTechnique obtains unique seeds for each query in Phase II (see
\cref{sec:guided-recommendation}), it combines them into a holistic configuration through a
\textit{rank}-and-\textit{compose} process. We discuss each part separately.

\subsection{Ranking Seeds}
\label{sec:ranking-seeds}

As a query's seeds can have diverse performance outcomes, ranking them is crucial
for composing into a holistic configuration. The simplest approach is to use the
DBMS's estimated plan cost. However, plan cost is not necessarily correlated with plan
quality~\cite{bergmann25}, particularly when query hints are involved that distort the cost.

\dbTechnique instead executes each seed to estimate its quality.
Deploying all seeds on the DBMS (e.g., building all indexes) induces significant overhead. Thus,
\dbTechnique derives alternate indexes that cover the original indexes~\cite{subotic18} and then
estimates each seed by substituting indexes in its original plan with an
alternate index that upper bounds the original plan's runtime.
For example, consider three seeds:
(1) \dbQuery{1} with \dbIndex{1}~\verb|t(a,b)|,
(2) \dbQuery{2} with \dbIndex{2}~\verb|t(a)| and \dbQuery{2} does not access \verb|b|, and
(3) \dbQuery{3} with \dbIndex{3}~\verb|t(a)| and \dbQuery{3} accesses \verb|b|.
For \dbQuery{1} and \dbQuery{2}, \dbTechnique builds \dbIndex{1} and forces both to use it
with their original plan. However, if \dbTechnique forces \dbQuery{3} to use \dbIndex{1},
then \dbQuery{3} may use the covering index to avoid heap fetches and execute faster than possible.
To avoid this, \dbTechnique builds \dbIndex{3} and forces \dbQuery{3} to use \dbIndex{3}.

\dbTechnique executes all seeds in order of increasing plan cost. If the DBMS
distorts the cost 
due to query hints (e.g., turn off sorting
when the query requires sorting), \dbTechnique computes a hypothetical cost without
distortions. To control query execution overhead, \dbTechnique applies a per-query timeout.

\subsection{Composition Algorithm}
\label{sec:composition-algorithm}

\begin{algorithm}[t!]
    {\footnotesize \begin{algorithmic}[1]
\State \textbf{Input:} Query-Configs $QC = \{\dots,(q_i, [c_{i1},\dots,c_{ij}])\}$
\State \textbf{Output:} Target Configuration

\State $seeds = \{(q_i, best(c_i)) \forall (q_i, c_i) \in QC\}$ \Comment{\circled{1}}
\State $cands = \texttt{\textbf{Merge}}(seeds)$ \Comment{\circled{2}}
\State $best = \texttt{\textbf{Evaluate}}(cands)$ \Comment{\circled{3}}
\While{$\text{budget not exhausted}$}
    \State $q_{next} = \texttt{\textbf{Select}}(best)$ \Comment{\circled{4}}
    \State $alt\_q\_seeds = \texttt{\textbf{Rollout}}(q_{next}, best, QC)$ \Comment{\circled{5}}
    \State $cands = \texttt{\textbf{Merge}}(best, q_{next}, alt\_q\_seeds)$ \Comment{\circled{6}}
    \State $best = \texttt{\textbf{Evaluate}}(cands)$
\EndWhile
\State \textbf{Return} $best$
\end{algorithmic}
}
    \caption{Beam Search}
    \label{alg:beam-search}
\end{algorithm}

\dbTechnique next composes a holistic configuration.
Inspired by relaxation-based physical design~\cite{bruno05relaxation},
\dbTechnique greedily constructs an initial configuration and then uses beam search
(i.e., best-first search~\cite{chen23loger,pearl84,azizi23elpis}) to refine further. In the
best case, combining each query's best seed results in a near-optimal configuration.
If conflicts arise (e.g., a query's runtime degrades from its seed's runtime),
\dbTechnique alters the configuration by targeting those conflicting seeds.

\cref{alg:beam-search} shows the beam search algorithm that runs until it exhausts the
time budget. \circled{1} \dbTechnique obtains the best seed
for each workload query, \circled{2} composes them
into holistic candidates (\textbf{Merge}), \circled{3} evaluates those candidates, and
selects the best one. From the best candidate, \circled{4} \dbTechnique
selects a query to refine (\textbf{Select}),
\circled{5} rolls out alternate seeds for the selected query (\textbf{Rollout}), and
\circled{6} repeats from \circled{2} by swapping the existing query seed out with alternate ones.
Our evaluation in \cref{sec:sensitivity-time} shows that composing for 1.5 hours produces good
results.
We next discuss these three steps.

\paragraph{Merge:}
In this step, \dbTechnique composes the seed for each query into a holistic configuration. As
the seeds may have incompatible parameters (e.g., different numbers of parallel workers),
\dbTechnique must reconcile those differences.
It only has to handle system knobs and physical design structures, since query knobs do not conflict
across queries. For system knobs,
\dbTechnique generates configurations based on the minimum, median, and maximum across the seeds.
For physical design structures, \dbTechnique takes the union and removes identical ones.
It then runs these holistic configurations on the DBMS 
with a cache to eliminate identical plan invocations~\cite{lim23}.

\paragraph{Select:}
\dbTechnique picks the next query to refine. \dbTechnique first fixes conflicts (i.e., degraded
queries) that result from merging into a holistic configuration. During this phase, \dbTechnique
greedily picks a query that performed worse than its estimated runtime. Afterwards, \dbTechnique
greedily picks the query with the highest runtime to refine next.
\dbTechnique tracks selected queries in query-agnostic configurations (i.e., only system knobs and
indexes) to avoid accidentally re-selecting a query \dbQuery{1} if another query \dbQuery{2} has
only changed its query hints that will not impact \dbQuery{1}'s plan.

\paragraph{Rollout:}
Last, \dbTechnique generates alternate seeds for the selected query. It then merges each alternate
seed with the other queries' seeds into holistic configurations and evaluates them. \dbTechnique
generates these alternate seeds with the following constrained exploration mechanisms
(\textbf{M1--4}) based on the query's present seed in the holistic configuration.
We analyze these further in \cref{sec:sensitivity-rollout}.

\begin{itemize}[leftmargin=*,topsep=2pt]
    \item
    \textbf{M1 Query Knob Permutation}: Similar to in Phase II
    (see \cref{sec:recommendation-sanitization}), this mechanism permutes the seed's query knobs
    to generate diverse plans (i.e., seeds). Common variations include deferring to the
    optimizer~\cite{chen23leon,zhang24holon} or forcing a nested loop join~\cite{marcus21,job}.

    \item
    \textbf{M2 Plan Repair}: This corrects the query's plan under the holistic
    configuration to match the seed's query plan. \dbTechnique analyzes both plans
    and then generates alternate seeds 
    by modifying query hints to restore or eliminate nodes.
    For example, if the holistic configuration's plan contains a \verb|Sort| node that is not
    in the seed's, then \dbTechnique creates an alternate seed with sorting disabled.
    \dbTechnique attempts three repairs for each query:
    (1) enable optimizer flags corresponding to used nodes,
    (2) turn off nodes not in the seed's plan, and
    (3) enable nodes in the seed's plan.

    \item
    \textbf{M3 Hidden Indexes}: Combining index sets together may impair
    cost-based query optimization~\cite{bruno05relaxation}. \dbTechnique first obtains the
    indexes used by the query under the holistic configuration. \dbTechnique then builds alternate
    seeds that omit those indexes, aiming to uncover alternative plans 
    that improve the query's performance.

    \item
    \textbf{M4 Ranked Seeds}: Each query has many ranked seeds that may compose with the other
    queries' seeds into different holistic configurations. \dbTechnique selects ranked seeds
    that could improve the query by selecting those with estimated runtimes less than the query's
    runtime under the holistic configuration.

\end{itemize}

During rollout, \dbTechnique may evaluate a large number (e.g., $\approx$100) of holistic
configurations. If an alternate seed only alters the query
knobs, then its changes to the holistic configuration are query-local (i.e., does not impact other
queries) and \dbTechnique uses per-query timeouts to bound the runtime. If the
seed alters the physical configuration, then \dbTechnique re-executes the entire workload.

%

Motivated by this, \dbTechnique adopts a 5-step rollout: 
(1) \textbf{Local},
(2) \textbf{25\% Seed},
(3) \textbf{50\% Seed},
(4) \textbf{75\% Seed}, and
(5) \textbf{100\% Seed}.
In \textbf{Local}, \dbTechnique uses \textbf{M1,M2} to try query-local alternative seeds
that are fast to evaluate. In the other steps, \dbTechnique explores
seeds from \textbf{M3,M4} in slices based on estimated runtime. Thus, 
\dbTechnique avoids suboptimal seeds if it discovers a performant one in an earlier step.


\section{Integration}
\label{sec:integration}

We now discuss how \dbTechnique integrates with existing tuners. As \dbTechnique is tuner-agnostic,
developers can plug it into 
a tuner through its API commands: (1) \dbAPI{Parse}, (2)
\dbAPI{Link}, and (3) \dbAPI{Digest}.

\paragraph{\textbf{\dbAPI{Parse}}}
As tuners generate artifacts in various formats,
this command enables \dbTechnique to decipher them and extract their key insights. This process
reconciles differences in tunables (e.g., knobs, query hints) between what a tuner supports
and what \dbTechnique reasons over. For example, some tuners only consider whether sequential scans
or index scans are enabled system-wide~\cite{zhang23unitune,wang21udo} or on a
per-query~\cite{marcus21,anneser23} basis.
By contrast, \dbTechnique reasons over access methods at a
finer table-level granularity for each query.

\paragraph{\textbf{\dbAPI{Link}}}
Developers implement how \dbTechnique links \dbQC{s}. Recall that the \dbQC
retrieval process uses these links to retrieve semantically relevant and performant \dbQC{s}
(see \cref{sec:guided-recommendation}). We propose linking based on the
temporal nature of tuning steps. By default, \dbTechnique links \dbQC objects based on explored
trajectories: \dbConfig{1} links to \dbConfig{2} if \dbConfig{2} is reached (i.e., explored)
from \dbConfig{1}.

\paragraph{\textbf{\dbAPI{Digest}}}
Lastly, once \dbTechnique completes its composition, \dbTechnique passes its
findings to the tuner for further refinement. By default, \dbTechnique exposes its best
holistic configuration 
to the 
tuner through \dbAPI{Digest}. However,
\dbTechnique could expose additional artifacts (e.g., explored configurations)
or guide further tuning focus (e.g., target specific queries). 
We envision that future tuners will support the ingestion of this data.
We defer this to future work.

\section{Evaluation}
\label{sec:eval}

We evaluate \dbTechnique's ability to accelerate the convergence of existing tuners when confronted
by different drifts and transfer scenarios. We target \pg v15.1 running on a
server with two Intel Xeon Gold 5218R CPUs (20 cores) and a 960~GB Samsung NVMe SSD. We restrict
the DBMS to 32~GB of RAM and 20 worker processes. To support \dbTechnique's operations in \pg,
we install \textit{\dbHypoPG}~\cite{hypopg} for what-if mechanism~\cite{surajit98autoadmin},
\textit{\dbPGHintPlan}~\cite{pg_hint} for query tuning, and a custom
\textit{\dbHypoExec} extension for re-costing query plans and swapping indexes during execution.

We primarily evaluate with three OLAP workloads. \textbf{JOB}~\cite{job} is a benchmark that
stresses the query optimizer with 21 tables and 113 queries. \textbf{TPC-H SF10}~\cite{tpch} models
a business analytics workload with eight tables and 22 queries. \textbf{DSB SF10}~\cite{dsb} is
Microsoft's extension of TPC-DS~\cite{tpcds} that introduces additional challenges (e.g., data
distributions, join patterns) with 25 tables and 53 queries. We omit four queries (\dbQuery{18},
\dbQuery{32}, \dbQuery{81}, \dbQuery{92}) due to \pg's query optimizer's limited ability to unnest
subqueries~\cite{franz23}.


\subsection{Experiment Setup}
\label{sec:eval-setup}
We deploy four state-of-the-art DBMS tuners:


\paragraph{PGTune+DTA+AutoSteer (\dbPDA)}
This first runs \dbPGTune~\cite{pgtune}, a heuristics-based knob tuner, followed by Microsoft's
Anytime Database Tuning Advisor (DTA) algorithm~\cite{surajit20anytime}. We use Hyrise's
implementation of \dbDTA~\cite{kossmann20} with an unlimited tuning budget. After \dbDTA finishes,
we run \dbAS~\cite{anneser23} to tune query knobs by greedily toggling and merging boolean
knobs.

\paragraph{\dbUniTune}
Alibaba's coordinating tuning framework that targets system
knobs, indexes, and limited query rewriting through Calcite~\cite{zhang24holon}. We adopt the same
settings as their
paper, but modify it to run queries serially and to minimize the workload runtime.

\paragraph{\dbPX}
CMU's holistic tuner that targets system
knobs, indexes, and query options supported by \dbPGHintPlan~\cite{zhang24holon}. We adopt the same
parameters from
their paper and extend it to suggest index or bitmap scans and common table expressions (CTEs)
inlining.

\paragraph{LambdaTune+AutoSteer (\dbLT)}
This first runs Cornell's LLM-based tuning agent (\dbLambda~\cite{giannakouris25}) that analyzes
the workload, constructs a tuning prompt based on the analysis, and obtains configurations
from \lambdagpt~\cite{gpt4o}. After obtaining a configuration, it then runs \dbAS~\cite{anneser23}
to tune query knobs.

\vspace*{0.05in}

\begin{figure*}[t!]
    \begin{subfigure}{.5\linewidth}
        \centering
        \setlength{\fboxsep}{0pt}
        \framebox[.95\linewidth]{
            \includegraphics[height=0.12in] {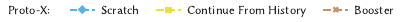}
        }
        \includegraphics[width=.95\linewidth]{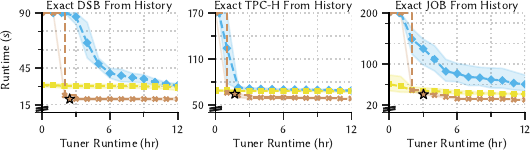}
        \captionsetup{aboveskip=0pt}
        \caption{\dbPX}
        \label{fig:exact-transfer-px}
    \end{subfigure}%
    \begin{subfigure}{.5\linewidth}
        \centering
        \setlength{\fboxsep}{0pt}
        \framebox[.95\linewidth]{
            \includegraphics[height=0.12in]{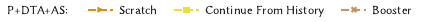}
        }
        \includegraphics[width=.95\linewidth]{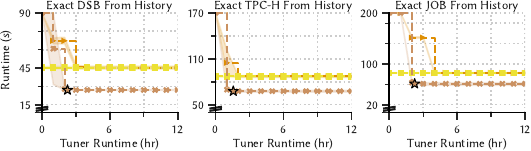}
        \captionsetup{aboveskip=0pt}
        \caption{\dbPDA}
        \label{fig:exact-transfer-pda}
    \end{subfigure}%

    \begin{subfigure}{.5\linewidth}
        \centering
        \setlength{\fboxsep}{0pt}
        \framebox[.95\linewidth]{
            \includegraphics[height=0.12in]{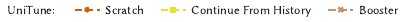}
        }
        \includegraphics[width=.95\linewidth]{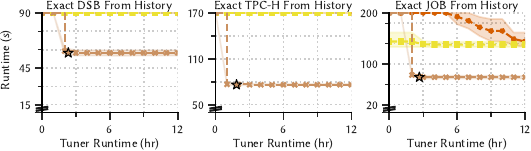}
        \captionsetup{aboveskip=0pt}
        \caption{\dbUniTune}
        \label{fig:exact-transfer-unitune}
    \end{subfigure}%
    \begin{subfigure}{.5\linewidth}
        \centering
        \setlength{\fboxsep}{0pt}
        \framebox[.95\linewidth]{
            \includegraphics[height=0.12in]{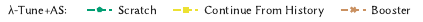}
        }
        \includegraphics[width=.95\linewidth]{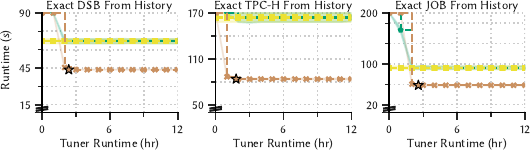}
        \captionsetup{aboveskip=0pt}
        \caption{\dbLT}
        \label{fig:exact-transfer-lt}
    \end{subfigure}

    \caption{
        \captionTitle{Exact Transfer}
        The DBMS's performance achieved by each framework on DSB, JOB, and TPC-H when tuning
        from scratch, from the best historical configuration, and accelerated with \dbTechnique.
        We plot the mean performance obtained by four trials of each tuner, with the error band
        representing the 95\% confidence interval. For \dbTechnique, the $\star$ illustrates when
        it finishes composing a holistic configuration.
    }
    \label{fig:exact-transfer}
\end{figure*}

We next briefly discuss \dbTechnique's standard configuration for all experiments. For local LLM
inference, \dbTechnique uses an RTX3080 GPU with 10GB of RAM, an Intel Xeon W1350 CPU (12 cores)
and 32GB of RAM. \dbTechnique uses \techniqueembed~\cite{voyageai} for its embedding model
to generate identity vectors (i.e., embeddings) and \techniqueprompt~\cite{llama31} to
suggest configurations. For consistency, we invoke the LLM with a temperature of $0$, a context
window of $16384$ tokens, and maximum output of $4096$ tokens.

In Phase I, we configure \dbTechnique to reason over configurations that span \dbPX's options:
system knobs, indexes, and query options for each query that cover optimizer switches, access method
selection (e.g., sequential, index, bitmap), and whether to materialize or inline
CTEs. In Phase II, \dbTechnique uses all permutation strategies to generate query seeds
(\cref{sec:recommendation-sanitization}). In Phase III, \dbTechnique runs its constrained
composition for 1.5hr (\cref{sec:composition-algorithm}).

We populate a tuning repository for each tuner by running four trials with
random seeds on the same hardware.
At the start of each trial, we initialize \pg with its default configuration.
We then run each tuner for 12hr to tune the DBMS. While these
tuners run,
\dbTechnique in Phase I (\cref{sec:qc-construction}) analyzes their artifacts and
generates \dbQC{s} that are available once the tuners finish.

For tuners \textit{continuing from history},
we start each tuner from one of the best configurations in the repository and allow it 12hr to 
optimize the DBMS further. When accelerating a tuner with \dbTechnique,
each \dbTechnique invocation runs Phase II and III with access to all artifacts from that
tuner's repository (e.g., four trials of \dbPX). After each tuner's trial, we re-evaluate 
discovered configurations without timeouts in a warm cache to obtain their actual
performance~\cite{zhang24holon,lehmann2024learned}.

\subsection{Exact Transfer}
\label{sec:eval-exact-transfer}

We first evaluate \dbTechnique's ability to accelerate tuners when tuning the same deployment from 
history. We compare three tuner variations: (1) tuning from scratch, (2) continuing from the best 
historical configuration, and (3) accelerated with \dbTechnique. We report each tuner's best
configuration mean performance, the 95\% confidence interval band, and the
$\star$ to indicate when \dbTechnique completes. We also show each tuner's worst, mean, and best
performance in \cref{tab:exact-transfer-tbl}.

As shown in \cref{fig:exact-transfer}, \textbf{Continue From History} is generally not an effective 
strategy. \dbPDA and \dbLT (\cref{fig:exact-transfer-pda,fig:exact-transfer-lt}) exhibit no 
improvement, whereas \dbUniTune (\cref{fig:exact-transfer-unitune}) achieves only 3--4\% mean 
improvement on TPC-H and JOB. \cref{tab:exact-transfer-tbl} indicates that although \dbPX achieves
only
1\% and 6\% improvement on TPC-H and DSB, respectively, its JOB configuration is 33\% better.

These results show \dbTechnique enables tuners to discover configurations that are better than
tuning from scratch or continuing from history. \dbTechnique helps to find
configurations that have a mean improvement of 16--51\% (\dbPX), 22--40\% (\dbDTA), 49--62\% 
(\dbUniTune), and 34--50\% (\dbLT) over tuning from scratch. \dbTechnique achieves this by
analyzing
artifacts for query-level insights and composing those insights together with
constrained exploration. It extends the set of tunables (e.g., query hints) supported by
the tuners 
and also breaks their search algorithms out of local optima.

\begin{table}[t!]
    \centering
    \caption{
        \captionTitle{Exact Transfer Performance Spread}
        The worst, mean, and best result for the tuners' trials in
        \cref{fig:exact-transfer} on DSB, TPC-H, and JOB when tuning from scratch,
        continuing from history, and with \dbTechnique.
    }
    \resizebox{\linewidth}{!}{

\begin{tabular}{@{}l|ccc|ccc|ccc@{}}
\toprule
    & \multicolumn{3}{c}{\textbf{Scratch}}
    & \multicolumn{3}{c}{\textbf{Continue}}
    & \multicolumn{3}{c}{\textbf{\dbTechnique}}
    \\
    Benchmark
    & \multicolumn{1}{c}{Min} & \multicolumn{1}{c}{Mean} & \multicolumn{1}{c}{Max}
    & \multicolumn{1}{c}{Min} & \multicolumn{1}{c}{Mean} & \multicolumn{1}{c}{Max}
    & \multicolumn{1}{c}{Min} & \multicolumn{1}{c}{Mean} & \multicolumn{1}{c}{Max}
    \\
\hline


\textbf{DSB $\!\rightarrow\!$ DSB} & & & & & & \\

\rowcolor{RowGray}
\hspace{3mm}\dbPX
& 29s & 31s & 33s
& 26s & 29s & 32s
& \textbf{19s} & \textbf{20s} & \textbf{21s}
\\
\hspace{3mm}\dbPDA
& 44s & 45s & 47s
& 44s & 45s & 47s
& \textbf{27s} & \textbf{27s} & \textbf{27s}
\\
\rowcolor{RowGray}
\hspace{3mm}\dbUniTune
& 125s & 136s & 148s
& 125s & 136s & 148s
& \textbf{55s} & \textbf{57s} & \textbf{60s}
\\
\hspace{3mm}\dbLT
& 64s & 67s & 70s
& 64s & 67s & 70s
& \textbf{44s} & \textbf{44s} & \textbf{44s}
\\
\hline
\textbf{TPC-H $\!\rightarrow\!$ TPC-H} & & & & & & \\

\rowcolor{RowGray}
\hspace{3mm}\dbPX
& 68s & 69s & 71s
& 66s & 68s & 69s
& \textbf{58s} & \textbf{58s} & \textbf{59s}
\\
\hspace{3mm}\dbPDA
& 86s & 87s & 88s
& 86s & 87s & 88s
& \textbf{66s} & \textbf{68s} & \textbf{69s}
\\
\rowcolor{RowGray}
\hspace{3mm}\dbUniTune
& 199s & 200s & 201s
& 176s & 194s & 200s
& \textbf{76s} & \textbf{76s} & \textbf{78s}
\\
\hspace{3mm}\dbLT
& 156s & 167s & 181s
& 156s & 167s & 181s
& \textbf{83s} & \textbf{84s} & \textbf{85s}
\\
\hline
\textbf{JOB $\!\rightarrow\!$ JOB} & & & & & & \\

\rowcolor{RowGray}
\hspace{3mm}\dbPX
& 47s & 61s & 88s
& 40s & 41s & 42s
& \textbf{29s} & \textbf{30s} & \textbf{32s}
\\
\hspace{3mm}\dbPDA
& 82s & 83s & 83s
& 82s & 83s & 83s
& \textbf{57s} & \textbf{61s} & \textbf{64s}
\\
\rowcolor{RowGray}
\hspace{3mm}\dbUniTune
& 132s & 144s & 170s
& 132s & 138s & 146s
& \textbf{72s} & \textbf{74s} & \textbf{78s}
\\
\hspace{3mm}\dbLT
& 89s & 93s & 97s
& 89s & 93s & 97s
& \textbf{58s} & \textbf{58s} & \textbf{60s}
\\
\hline
\end{tabular}
}
    \label{tab:exact-transfer-tbl}
\end{table}

\subsection{Parameter Drift}
\label{sec:eval-parameter-drift}

\begin{figure*}[t!]
    \begin{subfigure}{.5\linewidth}
        \centering
        \setlength{\fboxsep}{0pt}
        \framebox[.95\linewidth]{\includegraphics[height=0.12in]{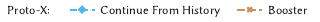}}

        \includegraphics[width=.95\linewidth]{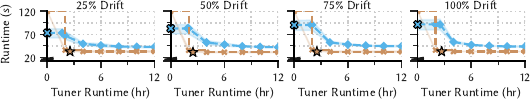}
        \captionsetup{aboveskip=0pt}
        \caption{\dbPX}
        \label{fig:pd-px}
    \end{subfigure}%
    \begin{subfigure}{.5\linewidth}
        \centering
        \setlength{\fboxsep}{0pt}
        \framebox[.95\linewidth]{\includegraphics[height=0.12in]{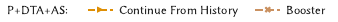}}

        \includegraphics[width=.95\linewidth]{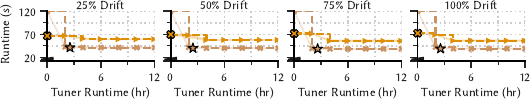}
        \captionsetup{aboveskip=0pt}
        \caption{\dbPDA}
        \label{fig:pd-pdtaas}
    \end{subfigure}
    
    \vspace{0.5em}
    \begin{subfigure}{.5\linewidth}
        \centering
        \setlength{\fboxsep}{0pt}
        \framebox[.95\linewidth]{\includegraphics[height=0.12in]{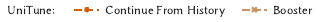}}

        \includegraphics[width=.95\linewidth]{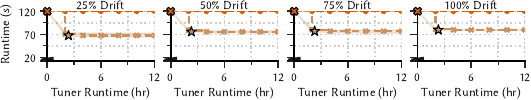}
        \captionsetup{aboveskip=0pt}
        \caption{\dbUniTune}
        \label{fig:pd-unitune}
    \end{subfigure}%
    \begin{subfigure}{.5\linewidth}
        \centering
        \setlength{\fboxsep}{0pt}
        \framebox[.95\linewidth]{\includegraphics[height=0.12in]{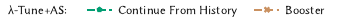}}

        \includegraphics[width=.95\linewidth]{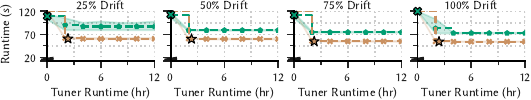}
        \captionsetup{aboveskip=0pt}
        \caption{\dbLT}
        \label{fig:pd-lambda}
    \end{subfigure}
    \caption{
        \captionTitle{Parameter Drift}
        The DBMS's performance achieved by each framework in response to workload parameter drift.
        We plot the mean performance with a 95\% confidence interval obtained by four trials of each
        tuner. The \textbf{\texttt{X}} represents the starting point when continuing from history. 
        For \dbTechnique,
        the $\star$ illustrates when it finishes composing a holistic configuration.
    }
    \label{fig:parameter-drift}
\end{figure*}

We next evaluate \dbTechnique's ability when the workload undergoes a parameter drift, whereby only 
the parameters of the queries change. We generate a set of DSB queries with a different seed and 
replace portions of the historical workload: \textbf{25\%}, \textbf{50\%}, \textbf{75\%}, and 
\textbf{100\%}. We run four 12hr trials for each drift percentage and plot the mean performance 
along with 95\% confidence interval in \cref{fig:parameter-drift}.

As shown in \cref{fig:parameter-drift} with the $\star$ marker, \dbTechnique
consistently outputs a holistic configuration in under 3hr. With \dbTechnique, all
frameworks find configurations with mean improvements of 23--27\% (\dbPX), 31\% (\dbPDA), 42--64\% 
(\dbUniTune), and 24--31\% (\dbLT) over those found by continuing to tune from historical
configurations. 
\dbTechnique re-mixes query seeds based on query semantics,
rather than the entire workload. For repeated queries, \dbTechnique 
extracts beneficial seeds ``as-is''. For queries with changed parameters, \dbTechnique generalizes 
from historical seeds with the same query template. Thus, it exploits the observation that
queries with the same templates may benefit from similar optimizations.
\dbTechnique assists in finding these configurations up to 3.6$\times$ (\dbPX), 
1.9$\times$ (\dbDTA), and 3.6$\times$ (\dbUniTune) faster. \dbLT finds configurations 10--30min 
faster than when using \dbTechnique. \dbTechnique spends $\sim$10min longer on LLM
inference, as it prompts the LLM on a query rather than workload granularity. Nevertheless, \dbLT 
with \dbTechnique still finds configurations that are 24--31\% better than without.

\subsection{Template Drift}
\label{sec:eval-template-drift}

\begin{figure*}[t!]
    \begin{subfigure}{.5\linewidth}
        \centering
        \setlength{\fboxsep}{0pt}
        \framebox[.95\linewidth]{\includegraphics[height=0.12in]{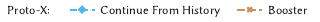}}

        \includegraphics[width=.95\linewidth]{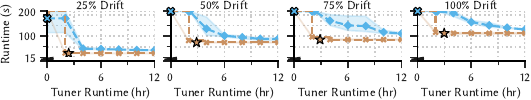}
        \captionsetup{aboveskip=0pt}
        \caption{\dbPX}
        \label{fig:td-px}
    \end{subfigure}%
    \begin{subfigure}{.5\linewidth}
        \centering
        \setlength{\fboxsep}{0pt}
        \framebox[.95\linewidth]{\includegraphics[height=0.12in]{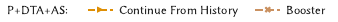}}

        \includegraphics[width=.95\linewidth]{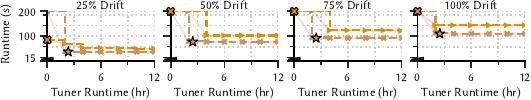}
        \captionsetup{aboveskip=0pt}
        \caption{\dbPDA}
        \label{fig:td-pdtaas}
    \end{subfigure}

    \vspace{0.5em}
    \begin{subfigure}{.5\linewidth}
        \centering
        \setlength{\fboxsep}{0pt}
        \framebox[.95\linewidth]{\includegraphics[height=0.12in]{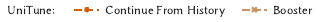}}

        \includegraphics[width=.95\linewidth]{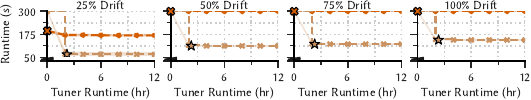}
        \captionsetup{aboveskip=0pt}
        \caption{\dbUniTune}
        \label{fig:td-unitune}
    \end{subfigure}%
    \begin{subfigure}{.5\linewidth}
        \centering
        \setlength{\fboxsep}{0pt}
        \framebox[.95\linewidth]{\includegraphics[height=0.12in]{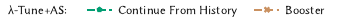}}

        \includegraphics[width=.95\linewidth]{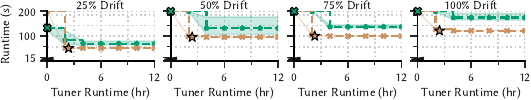}
        \captionsetup{aboveskip=0pt}
        \caption{\dbLT}
        \label{fig:td-lambda}
    \end{subfigure}
    \caption{
        \captionTitle{Template Drift}
        The DBMS's performance achieved by each framework in response to workload template drift.
        We plot the mean performance with a 95\% confidence interval obtained by four trials of each
        tuner. The \textbf{\texttt{X}} represents the starting point when continuing from history. 
        For \dbTechnique,
        the $\star$ illustrates when it finishes composing a holistic configuration.
    }
    \label{fig:template-drift}
\end{figure*}

We now assess how \dbTechnique manages workloads that experience template drifts. We 
generate TPC-DS~\cite{tpcds} queries that are not also DSB templates and replace portions of 
the historical workload by \textbf{25\%}, \textbf{50\%}, \textbf{75\%}, and \textbf{100\%}. We 
run four 12hr trials per percentage and plot the mean performance along with 95\% 
confidence interval.

Similar to the previous experiment, \cref{fig:template-drift} shows that
\dbTechnique 
outputs a holistic configuration in
under three hours ($\star$ marker). \dbTechnique enables 
the tuners to find configurations with mean improvements of 12--30\% (\dbPX),
24\% (\dbPDA), 52--63\% (\dbUniTune), and 26--39\% (\dbLT) over those found by tuning from
historical configurations.
\dbTechnique assists in finding these configurations up to
4.7$\times$ (\dbPX), 3.4$\times$ (\dbPDA), 2.7$\times$ (\dbUniTune), and 1.2$\times$ (\dbLT) 
faster. \dbTechnique handles unseen queries through two strategies. It first generalizes from 
similar historical queries. In
cases where the LLM generates a locally suboptimal configuration, \dbTechnique then relies on its
constrained exploration mechanism to derive similar
configurations that are more performant (\cref{sec:recommendation-sanitization}).

\subsection{Machine Transfer}
\label{sec:eval-machine-transfer}


\begin{figure}[t!]
    \begin{subfigure}{.5\linewidth}
        \centering
        \setlength{\fboxsep}{0pt}
        \framebox[.95\linewidth]{\includegraphics[height=0.12in]{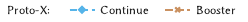}}

        \includegraphics[width=.95\linewidth]{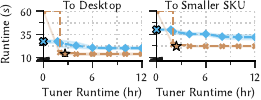}
        \captionsetup{aboveskip=0pt}
        \caption{\dbPX}
        \label{fig:mt-px}
    \end{subfigure}%
    \begin{subfigure}{.5\linewidth}
        \centering
        \setlength{\fboxsep}{0pt}
        \framebox[.95\linewidth]{\includegraphics[height=0.12in]{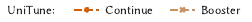}}

        \includegraphics[width=.95\linewidth]{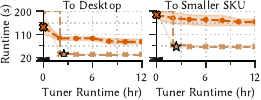}
        \captionsetup{aboveskip=0pt}
        \caption{\dbUniTune}
        \label{fig:mt-unitune}
    \end{subfigure}
    \caption{
        \captionTitle{Machine Transfer}
        The DBMS's performance achieved by each framework in response to machine transfer.
        We plot the mean performance with 95\% confidence interval obtained by four trials of each
        tuner. The \textbf{\texttt{X}} represents the starting point when continuing from history. 
        For \dbTechnique,
        the $\star$ illustrates when it finishes composing a holistic configuration.
    }
    \label{fig:machine-transfer}
\end{figure}

We next evaluate \dbTechnique's ability when the environment undergoes a hardware change.
We evaluate two specific scenarios moving from a high-performance server to a 
\textit{weaker} (Intel Xeon Silver 4114) and \textit{desktop} (Intel Xeon W-1350) machine with the 
same DSB
workload from history. We evaluate the best- and worst-performing tuners from
\cref{sec:eval-exact-transfer} using four 12hr trials for each environment.

As shown in \cref{fig:machine-transfer}, all tuners with \dbTechnique find
configurations with mean improvements of 31--32\% (\dbPX) and 61--62\% (\dbUniTune) over those found
by tuning
from historical configurations. Furthermore, \dbTechnique enables tuners to find
these configurations up to 3.1$\times$ (\dbPX) and 2.2$\times$ (\dbUniTune) faster. As the target 
workload is the
same as the history, \dbTechnique recognizes that each query's best historical seed is a promising
starting point. It then leverages this to assist both tuners in finding
better configurations more quickly.

\subsection{Dataset Growth}
\label{sec:eval-dataset-growth}

\begin{figure}[t!]
    \begin{subfigure}{.5\linewidth}
        \centering
        \setlength{\fboxsep}{0pt}
        \framebox[.95\linewidth]{\includegraphics[height=0.12in]{diagrams/volume/legend4.pdf}}

        \includegraphics[width=.95\linewidth]{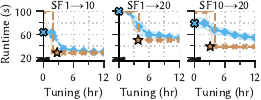}
        \captionsetup{aboveskip=0pt}
        \caption{\dbPX}
        \label{fig:dt-px}
    \end{subfigure}%
    \begin{subfigure}{.5\linewidth}
        \centering
        \setlength{\fboxsep}{0pt}
        \framebox[.95\linewidth]{\includegraphics[height=0.12in]{diagrams/volume/legend2.pdf}}

        \includegraphics[width=.95\linewidth]{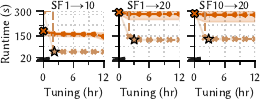}
        \captionsetup{aboveskip=0pt}
        \caption{\dbUniTune}
        \label{fig:dt-unitune}
    \end{subfigure}
    \caption{
        \captionTitle{Dataset Growth}
        The DBMS's performance achieved by each framework in response to dataset growth.
        We plot the mean performance with a 95\% confidence interval obtained by four trials of each
        tuner. The \textbf{\texttt{X}} represents the starting point when continuing from history. 
        For \dbTechnique,
        the $\star$ illustrates when it finishes composing a holistic configuration.
    }
    \label{fig:dataset-growth}
\end{figure}

\begin{figure*}[t!]
    \begin{subfigure}{.5\linewidth}
        \centering
        \setlength{\fboxsep}{0pt}
        \framebox[.95\linewidth]{\includegraphics[height=0.10in]{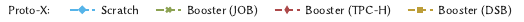}}
        \includegraphics[width=.95\linewidth]{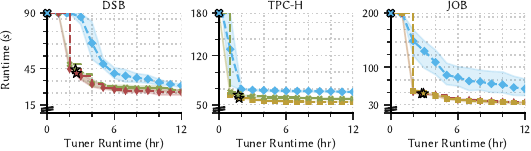}

        \captionsetup{aboveskip=0pt}
        \caption{\dbPX}
        \label{fig:cross-schema-px}
    \end{subfigure}%
    \begin{subfigure}{.5\linewidth}
        \centering
        \setlength{\fboxsep}{0pt}
        \framebox[.95\linewidth]{\includegraphics[height=0.10in]{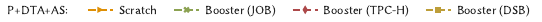}}
        \includegraphics[width=.95\linewidth]{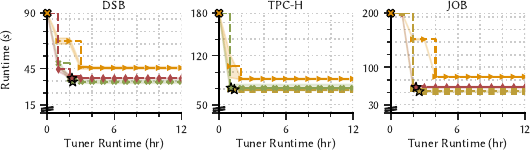}

        \captionsetup{aboveskip=0pt}
        \caption{\dbPDA}
        \label{fig:cross-schema-pda}
    \end{subfigure}

    \begin{subfigure}{.5\linewidth}
        \centering
        \setlength{\fboxsep}{0pt}
        \framebox[.95\linewidth]{\includegraphics[height=0.10in]{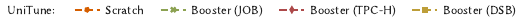}}
        \includegraphics[width=.95\linewidth]{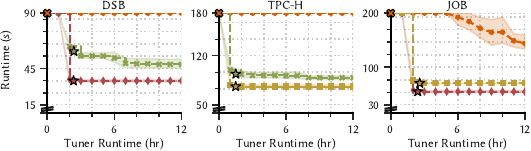}

        \captionsetup{aboveskip=0pt}
        \caption{\dbUniTune}
        \label{fig:cross-schema-unitune}
    \end{subfigure}%
    \begin{subfigure}{.5\linewidth}
        \centering
        \setlength{\fboxsep}{0pt}
        \framebox[.95\linewidth]{\includegraphics[height=0.10in]{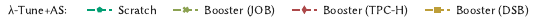}}
        \includegraphics[width=.95\linewidth]{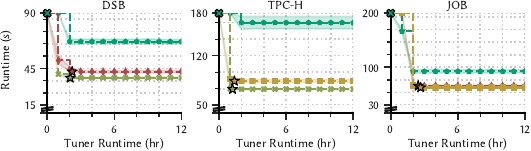}

        \captionsetup{aboveskip=0pt}
        \caption{\dbLT}
        \label{fig:cross-schema-lt}
    \end{subfigure}

    \caption{
        \captionTitle{Cross-Schema Transfer}
        The DBMS's performance achieved by each framework on DSB, JOB, and TPC-H when tuning
        from scratch and accelerated with \dbTechnique from other benchmarks' artifacts.
        We plot the mean performance obtained by four trials of each tuner, with the error band
        representing the 95\% confidence interval. For \dbTechnique, the $\star$ illustrates when
        it finishes composing a holistic configuration.
    }
    \label{fig:cross-schema}
\end{figure*}

\begin{figure}[t!]
    \centering
    \setlength{\fboxsep}{0pt}
    \framebox[\linewidth]{\includegraphics[height=0.12in]{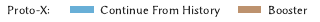}}

   \includegraphics[width=0.95\columnwidth]{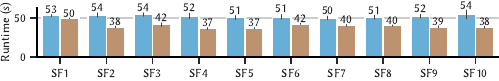}
   \caption{
       \captionTitle{Dataset Growth Sensitivity}
       Mean performance with 95\% confidence interval
       obtained by four trials of \dbPX. Evaluates the dataset growth scenario from different
       DSB SF to SF20.
   }
   \label{fig:dataset-growth-sens}
\end{figure}

Another challenge for tuners is when the database grows over time or when switching from
development to production instances. To measure this effect, we generate three DSB 
variations: \textbf{SF1$\rightarrow$10}, \textbf{SF1$\rightarrow$20}, and 
\textbf{SF10$\rightarrow$20}. Following the process 
in \cref{sec:eval-setup},
we first construct the artifacts repository from tuning \textbf{SF1}. We then evaluate the best- and
worst-performing frameworks from \cref{sec:eval-exact-transfer} using the same workload for four
12hr trials per scenario.

\cref{fig:dataset-growth}'s $\star$ marker indicates that \dbTechnique takes longer on 
\textbf{SF20}
due to the higher query execution overhead. As shown in \cref{fig:dt-unitune}, \dbTechnique
assists \dbUniTune in finding configurations with mean improvements of 63\% 
(\textbf{SF1$\rightarrow$10}),
55\% (\textbf{SF1$\rightarrow$20}), and 53\% (\textbf{SF10$\rightarrow$20}) over those found by 
tuning from
historical configurations.
 
In contrast in \cref{fig:dt-px}, \dbPX shows more modest improvements of 6\% 
(\textbf{SF1$\rightarrow$10}),
7\% (\textbf{SF1$\rightarrow$20}), and 29\% (\textbf{SF10$\rightarrow$20}). To investigate the 
performance
difference between adapting to \textbf{SF20} from \textbf{SF1} versus \textbf{SF10}, we sweep each 
SF
from 1 to 10 and evaluate with the same methodology as above. We run four 12hr trials for each
scenario and plot the mean performance along with 95\% confidence interval in
\cref{fig:dataset-growth-sens}. As \cref{fig:dataset-growth-sens} shows, adapting from SF1 stands
out, with the others resulting in configurations that are 18--30\% better.
Upon closer inspection of SF1's
artifacts generated by \dbPX, we notice several performant plans at SF1 size that do not
transfer to SF20. We defer building tuners with an explicit generalizability objective, in addition
to a performance objective, for future work.

\subsection{Cross-Schema Transfer}
\label{sec:eval-cross-schema}

\begin{table}[t!]
    \centering
    \caption{
        \captionTitle{Cross-Schema Transfer Performance Spread}
        The worst, mean, and best performance achieved by a framework's four trials in
        \cref{fig:cross-schema} on DSB, TPC-H, and JOB when tuning from scratch,
        and with \dbTechnique 
        from other benchmarks'
        repository artifacts.
    }
    \resizebox{\linewidth}{!}{\begin{tabular}{@{}l|ccc|ccc|ccc|ccc|@{}}
\toprule
    & \multicolumn{3}{c}{\textbf{Scratch}} 
    & \multicolumn{3}{c}{\textbf{\dbTechnique(DSB)}}
    & \multicolumn{3}{c}{\textbf{\dbTechnique(TPC-H)}} 
    & \multicolumn{3}{c}{\textbf{\dbTechnique(JOB)}}
    \\
    Benchmark 
    & \multicolumn{1}{c}{Min} & \multicolumn{1}{c}{Mean} & \multicolumn{1}{c}{Max}
    & \multicolumn{1}{c}{Min} & \multicolumn{1}{c}{Mean} & \multicolumn{1}{c}{Max}
    & \multicolumn{1}{c}{Min} & \multicolumn{1}{c}{Mean} & \multicolumn{1}{c}{Max} 
    & \multicolumn{1}{c}{Min} & \multicolumn{1}{c}{Mean} & \multicolumn{1}{c}{Max}
    \\
\hline
\textbf{DSB} & & & & & & & & & & & & \\

\rowcolor{RowGray}
\hspace{3mm}\dbPX
& 29s & 31s & 33s
& - & - & -
& 22s & \textbf{26s} & 28s
& 25s & \textbf{26s} & 27s
\\
\hspace{3mm}\dbPDA
& 44s & 45s & 47s
& - & - & -
& 36s & 37s & 37s
& 32s & \textbf{34s} & 36s
\\
\rowcolor{RowGray}
\hspace{3mm}\dbUniTune
& 125s & 136s & 148s
& - & - & -
& 34s & \textbf{35s} & 36s
& 44s & 48s & 55s
\\
\hspace{3mm}\dbLT
& 64s & 67s & 70s
& - & - & -
& 41s & 42s & 44s
& 36s & \textbf{37s} & 38s
\\
\hline
\textbf{TPC-H} & & & & & & & & & & & & \\

\rowcolor{RowGray}
\hspace{3mm}\dbPX
& 68s & 69s & 71s
& 54s & \textbf{55s} & 56s
& - & - & -
& 58s & 59s & 61s
\\
\hspace{3mm}\dbPDA
& 86s & 87s & 88s
& 70s & \textbf{72s} & 74s
& - & - & -
& 71s & 74s & 82s
\\
\rowcolor{RowGray}
\hspace{3mm}\dbUniTune
& 199s & 200s & 201s
& 74s & \textbf{76s} & 80s
& - & - & -
& 85s & 89s & 92s
\\
\hspace{3mm}\dbLT
& 156s & 167s & 181s
& 82s & 84s & 88s
& - & - & -
& 72s & \textbf{73s} & 73s
\\
\hline
\textbf{JOB} & & & & & & & & & & & & \\

\rowcolor{RowGray}
\hspace{3mm}\dbPX
& 47s & 61s & 88s
& 30s & \textbf{34s} & 37s
& 34s & 35s & 35s
& - & - & -
\\
\hspace{3mm}\dbPDA
& 82s & 83s & 83s
& 54s & \textbf{55s} & 57s
& 62s & 63s & 65s
& - & - & -
\\
\rowcolor{RowGray}
\hspace{3mm}\dbUniTune
& 132s & 144s & 170s
& 69s & 71s & 72s
& 53s & \textbf{55s} & 58s
& - & - & -
\\
\hspace{3mm}\dbLT
& 89s & 93s & 97s
& 61s & \textbf{63s} & 64s
& 64s & 66s & 67s
& - & - & -
\\
\hline
\end{tabular}
}
    \label{tab:cross-schema-tbl}
\end{table}

We next evaluate \dbTechnique's ability to accelerate tuners when tuning a new DBMS using artifacts
from a different application's database.
We run four trials
for each tuner variation across all workloads. We again plot the mean performance of each tuner's
best configuration, the 95\% confidence interval band, and use $\star$ to indicate when \dbTechnique
completes. We also report the worst, mean, and best performance achieved by any framework's trial in
\cref{tab:cross-schema-tbl}.

As indicated in \cref{fig:cross-schema}, tuning from the best historical
configuration is identical to tuning from scratch. Due to schema differences, existing tuning
frameworks cannot transfer the historical configurations. By contrast, \dbTechnique obtains
holistic configurations for the target DBMS that are inspired by artifacts, with configurations
that leverage \verb|INCLUDE| columns
(i.e., covering index) and index knobs (e.g., page fillfactor).
Thus, \dbTechnique assists tuners in discovering configurations with a mean improvement of 14--44\%
(\dbPX), 15--33\% (\dbPDA), 51--74\% (\dbUniTune), and 29--56\%
(\dbLT) over starting with the best historical configuration.
\cref{tab:cross-schema-tbl} shows that there is no dominant benchmark (i.e., artifact repository) to
initialize \dbTechnique with. Instead, \dbTechnique can use diverse artifacts to accelerate
tuners in finding better configurations.

\section{Sensitivity Experiments}
\label{sec:eval-sensitivity}

We next analyze aspects of \dbTechnique in more detail. We begin with an ablation study on
fine-tuning in \cref{sec:sensitivity-ft},
query knob permutation strategy (\cref{sec:recommendation-sanitization}) in
\cref{sec:sensitivity-dk},
the composition phase's search time in \cref{sec:sensitivity-time},
the embedder and prompter LLMs in \cref{sec:sensitivity-models},
the composition phase's rollout policy (\cref{sec:composition-algorithm}) in
\cref{sec:sensitivity-rollout},
and size of the input data in \cref{sec:sensitivity-input-data}.

\subsection{Fine-Tuning}
\label{sec:sensitivity-ft}

\begin{figure}[t!]
    \centering
    \setlength{\fboxsep}{0pt}
    \framebox[\linewidth]{\includegraphics[height=0.12in]{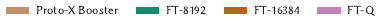}}
    \includegraphics[width=\linewidth]{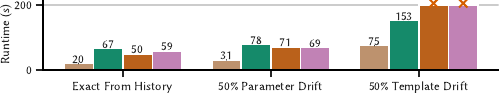}
    \caption{
        \captionTitle{Fine-Tuning}
        The DBMS's performance achieved by each technique across three scenarios
        based on \dbPX's historical DSB tuning artifacts. We plot the mean performance with a 95\%
        confidence interval obtained from four trials of each technique \textit{without}
        further refinement.
    }
    \label{fig:prompt-v-ft-sens}
\end{figure}

An alternative to \dbTechnique's approach of enriching the prompt with historical references
is fine-tuning an LLM on (prompt, configuration) pairs~\cite{huang2025e2e}.
We study whether fine-tuning alone is sufficient.
Using the most performant tuner \dbPX's DSB artifacts in the transfer, parameter, and template
experiments (\cref{sec:eval-exact-transfer,sec:eval-parameter-drift,sec:eval-template-drift}),
we extract each trial's steps into workload-level training data pairs.
For each step, we set the output to the trial's best configuration and
construct a prompt that contains tuning instructions, a workload summary, DBMS
metrics, and query plans truncated to specific context lengths
(i.e., \textbf{FT-8192}, \textbf{FT-16384}).
We explore a query variant (\textbf{FT-Q}) that replaces the workload summary with the SQL query.

We fine-tune \dbTechnique's LLM (\techniquepromptbone) with
LLaMa-Factory~\cite{zheng24llamafactory}, FlashAttention-2~\cite{dao2022flash}, and
DeepSpeed~\cite{rasley20deepspeed}. For all, we use a learning rate of 2e-5 for $8$
epochs~\cite{huang2025e2e}. We use 4 RTX3090 for \textbf{FT-8192} and \textbf{FT-Q} and
2 RTXA6000 for \textbf{FT-16384}. During inference for \textbf{FT-8192} and \textbf{FT-16384},
we sample 8 configurations at a temperature of 0.2~\cite{huang2025e2e} and select the best.
For \textbf{FT-Q}, we sample three configurations for each query with a temperature of 0.2, select 
each query's best config, and then combine them together. 

We consider three scenarios from the same historical DSB workload: \textbf{Exact From History},
\textbf{50\% Parameter Drift}, and \textbf{50\% Template Drift}. 
We run four trials for each technique \textit{without} further refinement. 
We plot the
mean performance along with a 95\% confidence interval in \cref{fig:prompt-v-ft-sens}.
Across all scenarios, fine-tuning performs worse. For \textbf{FT-8192} and \textbf{FT-16384},
truncating to context lengths restricts learning to the prompt's queries. For \textbf{FT-Q},
query configs conflict when combined, as seen in \textbf{50\% Template Drift}.

\subsection{Query Knob Permutation Strategy}
\label{sec:sensitivity-dk}

\dbTechnique permutes query knobs to generate similar query seeds 
(\cref{sec:recommendation-sanitization}).
We consider five strategies:
\textbf{None}, 
\textbf{Join Types} 
(i.e., hash, merge, nested loop),
\textbf{Access Methods} 
(i.e., seq-, bitmap-, or index-scan per-table),
\textbf{Sort} turns off sorting, 
and \textbf{All}. 
We limit the study to the most performant tuner \dbPX and evaluate three scenarios from the same
historical DSB workload: \textbf{Exact From History}, \textbf{50\% Parameter Drift}, and
\textbf{50\% Template Drift}. We run four trials for each strategy \textit{without} further
refinement by \dbPX. We plot the mean performance along with a 95\% confidence interval.

\begin{figure}[t!]
    \centering
    \setlength{\fboxsep}{0pt}
    \framebox[\linewidth]{\includegraphics[height=0.12in]{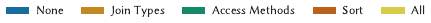}}

    \begin{subfigure}{\linewidth}
        \centering
        \includegraphics[width=\linewidth]{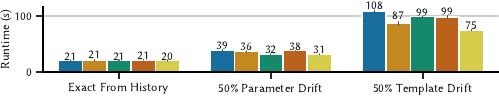}
    \end{subfigure}


    \caption{
        \captionTitle{Query Knob Permutation Strategy}
        The DBMS's performance achieved across three scenarios
        based on \dbPX's historical DSB tuning artifacts
        when varying \dbTechnique's query knob permutation strategy.
        We plot the mean performance with a 95\% confidence interval obtained from four trials of
        each
        technique \textit{without} further refinement.
    }
    \label{fig:exploration-switches}
\end{figure}

As shown in \cref{fig:exploration-switches}, the strategy does not matter for
\textbf{Exact From History}, as \dbTechnique exploits the 1:1 mapping
from the target workload to historical query seeds. For drifts, \textbf{Join Types},
\textbf{Access Methods}, and \textbf{Sort} strategies enable \dbTechnique to improve over
\textbf{None} by
allowing it to locally explore during composition to fix conflicts (i.e., degraded
queries) and break out of suboptimal solutions. 
\dbTechnique with the \textbf{All} strategy finds the best holistic configuration, as it uses all
opportunities to derive diverse plans from each query seed.

\subsection{Search Time}
\label{sec:sensitivity-time}

\begin{figure}[t!]
    \centering
    \setlength{\fboxsep}{0pt}
    \framebox[\linewidth]{\includegraphics[height=0.12in]{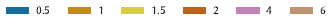}}

    \begin{subfigure}{\linewidth}
        \centering
        \includegraphics[width=\linewidth]{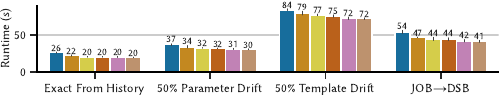}
    \end{subfigure}


    \caption{
        \captionTitle{Search Time}
        The DBMS's performance achieved across four scenarios
        based on \dbPX's historical DSB tuning artifacts
        when varying \dbTechnique's search time. 
        We plot the mean performance with a 95\% confidence interval obtained from four trials of
        each
        variant \textit{without} further refinement.
    }
    \label{fig:search-time}
\end{figure}

We next vary the amount of search time allocated to \dbTechnique's constrained composition
algorithm (\cref{sec:composition-algorithm}):
0.5hr, 1hr, 1.5hrs, 2hrs, 4hrs, and 6hrs. We limit the study to the most performant tuner \dbPX
and evaluate four scenarios from the same historical DSB workload: \textbf{Exact From History},
\textbf{50\% Parameter Drift}, \textbf{50\% Template Drift}, and \textbf{JOB$\rightarrow$DSB}.
We run four trials for each variation \textit{without} further refinement by \dbPX.
We report the mean performance along with a 95\% confidence interval. 

\cref{fig:search-time} shows that the trend across all four scenarios aligns with common tuning
trends~\cite{vanaken17,zhang24holon}. Providing \dbTechnique more time allows it to provide
better holistic configurations to the assisted tuner, with diminishing returns. Although
\textbf{Exact From History} stabilizes after 1.5hrs, the other scenarios continue to show
marginal improvement.

\subsection{Embedder-Prompter Models}
\label{sec:sensitivity-models}

\begin{figure}[t!]
    \begin{mdframed}[innertopmargin=0,innerbottommargin=0]
        \begin{subfigure}{.34\linewidth}            
            \includegraphics[height=0.11in]{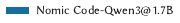}

            \vspace{-3pt}
            \includegraphics[height=0.11in]{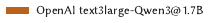}

            \vspace{-3pt}
            \includegraphics[height=0.11in]{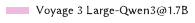}
        \end{subfigure}%
        \begin{subfigure}{.38\linewidth}            
            \includegraphics[height=0.11in]{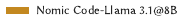}

            \vspace{-3pt}
            \includegraphics[height=0.11in]{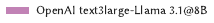}

            \vspace{-3pt}
            \includegraphics[height=0.11in]{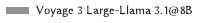}
        \end{subfigure}%
        \begin{subfigure}{.28\linewidth}            
            \includegraphics[height=0.11in]{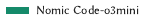}

            \vspace{-3pt}
            \includegraphics[height=0.11in]{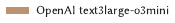}

            \vspace{-3pt}
            \includegraphics[height=0.11in]{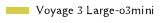}
        \end{subfigure}

    \end{mdframed}

    \begin{subfigure}{\linewidth}
        \centering
        \includegraphics[width=\linewidth]{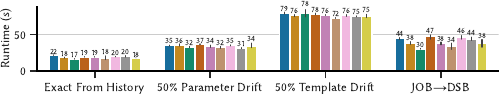}
    \end{subfigure}


    \caption{
        \captionTitle{Embedder-Prompter Models}
        The DBMS's performance achieved across four scenarios
        based on \dbPX's historical DSB tuning artifacts
        when varying the embedder and prompter LLM.
        We plot the mean performance with a 95\% confidence interval obtained from four trials of
        each
        variant \textit{without} further refinement.
    }
    \label{fig:embed-prompt}
\end{figure}

We next investigate the embedder and prompter LLMs used by \dbTechnique.
We select three embedders 
based on the Massive Text Embedding Benchmark~\cite{muennighoff2023mteb}:
\nomiccode~\cite{suresh2025nomic},
\textlarge~\cite{text3large},
and \techniqueembed~\cite{voyageai}.
We then select three prompters of different scales:
\qwen~\cite{yang2025qwen3technicalreport},
\techniquepromptbone~\cite{llama31}, 
and \othreemini~\cite{o3mini}.
We limit the study to the most performant tuner \dbPX and evaluate four scenarios from the same
historical DSB workload: \textbf{Exact From History}, \textbf{50\% Parameter Drift}, \textbf{50\%
Template Drift}, and \textbf{JOB$\rightarrow$DSB}.
We run four trials for each 
pair \textit{without} further refinement by \dbPX.

Examining \cref{fig:embed-prompt}, 
we first notice that the embedder-prompter matters more when the target workload differs from
historical workloads.
All pairs are comparable ($\approx$5s) for \textbf{Exact} but have larger differences for
\textbf{JOB$\rightarrow$DSB} ($\approx$17s).
Second, increasing the complexity of the prompter LLM
(i.e., \# of parameters)
enables \dbTechnique to find better configurations.
For instance, \dbTechnique with \othreemini finds configurations that are 32\% better than using
\qwen for \textbf{JOB$\rightarrow$DSB}. However, these improvements come with higher inference costs
that are not necessary for some scenarios (\textbf{Exact}, \textbf{50\% Parameter Drift}). We
leave the selection of the optimal LLM based on generalizability (e.g., cross-schema) and cost
requirements for future work~\cite{piskala2024optiroute}.

\subsection{Rollout Policy}
\label{sec:sensitivity-rollout}

\begin{figure}
    \centering
    \setlength{\fboxsep}{0pt}
    \framebox[\linewidth]{\includegraphics[height=0.12in]{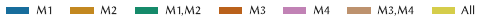}}
    \includegraphics[width=\linewidth]{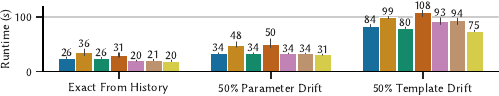}

    \caption{
        \captionTitle{Rollout Policy}
        The DBMS's performance achieved across three scenarios
        based on \dbPX's historical DSB tuning artifacts
        when varying \dbTechnique's rollout policy during composition.
        We plot the mean performance with a 95\% confidence interval obtained from four trials of
        each
        variant \textit{without} further refinement.
    }
    \label{fig:sensitivity-rollout}
\end{figure}

We next study \dbTechnique's rollout policy during composition
(\cref{sec:composition-algorithm}) and consider the following variations:
\textbf{M1 Query Knob Permutation},
\textbf{M2 Plan Repair},
\textbf{M3 Hidden Indexes},
\textbf{M4 Ranked Seeds},
\textbf{M1,M2} (i.e., query-local changes),
\textbf{M3,M4} (i.e., physical changes),
and \textbf{All}.
We limit the study to the most performant tuner \dbPX and
evaluate three scenarios from the same historical DSB workload: \textbf{Exact From History},
\textbf{50\% Parameter Drift}, and \textbf{50\% Template Drift}.
We run four trials for each variation \textit{without} further refinement by \dbPX. We plot the
mean performance along with a 95\% confidence interval. 

As shown in \cref{fig:sensitivity-rollout}, we find that \textbf{M2} and \textbf{M3} alone have
limited effectiveness due to their limited exploration. \textbf{M1} and \textbf{M4} are more
effective by providing a greater degree of diverse composition opportunities. We also observe that
although \textbf{M3,M4} (i.e., physical changes) finds better configurations in \textbf{Exact},
\textbf{M1,M2} (i.e., query-local changes) outperforms in \textbf{50\% Template Drift}.
\dbTechnique with \textbf{All} outputs the best configuration by leveraging all opportunities.

\subsection{Input Data}
\label{sec:sensitivity-input-data}

\begin{figure}
    \centering
    \setlength{\fboxsep}{0pt}
    \framebox[\linewidth]{\includegraphics[height=0.12in]{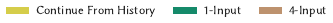}}

    \begin{subfigure}{\linewidth}
        \centering
        \includegraphics[width=\linewidth]{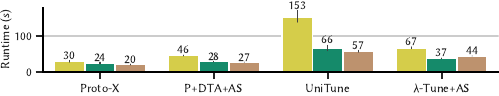}
        \captionsetup{aboveskip=0pt}
        \caption{Exact From History}
    \end{subfigure}



    \caption{
        \captionTitle{Input Data}
        The DBMS's performance achieved by different tuners when transferring the same workload
        as history. We evaluate whether \dbTechnique has access to one artifact or all four
        artifacts from each tuner. We plot the mean performance with a 95\% confidence interval
        obtained by four trials of each technique \textit{without} further refinement.
    }
    \label{fig:sensitivity-input}
\end{figure}

Recall from \cref{sec:eval-setup} that \dbTechnique ingests artifacts from all four tuner trials.
We next measure the efficacy of \dbTechnique when only using
one artifact
or all four artifacts for each tuner. We run four trials for each variation \textit{without}
further refinement. We plot the mean performance along with a 95\% confidence interval.

As shown in \cref{fig:sensitivity-input}, 
\dbTechnique with \textbf{1-Input} performs worse than the \textbf{4-Input} for \dbPX and
\dbUniTune.
By contrast, \dbTechnique with \textbf{1-Input} performs better for \dbLT. In the case of
\dbPX and \dbUniTune, providing \dbTechnique with more data allows it to leverage query-level
insights across a broader scope and avoid being overly restricted by a single trial's local optima.
For \dbLT, we find that \dbTechnique with \textbf{1-Input} augments its prompt with reference
\dbQC{s} from more diverse query templates (\cref{sec:guided-recommendation}) that lead to
more performant outcomes. We defer selecting the number of \dbQC{s} to enrich the
prompt based on each query to future work.

\section{Related Work}
\label{sec:related-work}

We now discuss existing autonomous DBMS research, focusing on
forecasting, behavior modeling, tuning frameworks, representations, natural language debugging, and
learned components.

\paragraph{Forecasting:}
Existing literature focuses on predicting future queries and loads for DBMS optimization.
These techniques include arrival rates~\cite{ma18qb}, parameterized queries~\cite{huang24sibyl},
and identifying ongoing workload drifts~\cite{li22warper,zhu23meter}.
Once these techniques identify a drift, a human operator gathers the new workload, target
DBMS deployment, and tuner and provides them to \dbTechnique to accelerate the tuner's adaptation
to the new environment.

\paragraph{Behavior Modeling:}
Modeling aims to produce accurate models for inferring the performance of the
DBMS~\cite{ma21mb2,shi22,lim24hit}. Behavior models can substitute for actual query
execution and for speculating on the impact of different configurations.
Furthermore, \dbTechnique generates substantial DBMS telemetry during operation, which a human
operator can use to create and refine these behavior models.

\paragraph{Tuning Frameworks:}
The literature is rich in tuning frameworks that target a single aspect of the DBMS:
resource-tuning~\cite{pgtune,vanaken17,kanellis22},
capacity planning~\cite{ammerlaan21,cahoon22},
physical design~\cite{dexter,zhou23,ahmed20,siddiqui22,wang24index,surajit20anytime},
and query tuning~\cite{marcus21,anneser23}.
Recent research has focused on tuning over multiple aspects of the
system~\cite{wang21udo,zhang23unitune,zhang24holon,giannakouris25}.

\paragraph{Representations:} This work focuses on deriving a query \cite{zhao22} or
workload~\cite{wangj23} representation that is conducive to downstream tasks (e.g.,
behavior models, tuning). These create models that output discriminative vector representations
(e.g., embeddings) based on inputs, such as query plans, workload properties, and DBMS statistics.
These could optionally be used by \dbTechnique to select relevant historical experiences.

\paragraph{Natural Language Debugging:}
With advances in LLMs, recent work has also focused on debugging SQL issues through a natural
language interface. These range from natural language to SQL techniques~\cite{fan24,zhao25},
root cause analysis~\cite{ouyang25rca},
and debugging user problems~\cite{wang25andromeda,zhou24dbot}.
These techniques primarily focus on designing retrieval mechanisms (e.g., ranking functions,
fine-tuned embedding models) for selecting the top-$k$ chunks to provide to the LLM based on the
user's question.

\paragraph{Learned Components:}
These are traditional DBMS subsystems augmented with machine learning.
Existing work has focused on layouts~\cite{ding21}, data structures~\cite{gu23},
query processing~\cite{sabek23}, and query optimization~\cite{marcus21,liy23,akioyamen24}.
Other research in this area has focused on learned cardinality estimation~\cite{negi23,wangf23},
due to its impact on query optimizer plan quality~\cite{lee23}.

\section{Conclusion}
\label{sec:conclusion}

Despite advances in tuners' ability to find performant configurations,
existing tuners remain unable to adapt to
environment changes (e.g., workload drifts, cross-schema transfers) due to their design.
To remedy this, we present the \dbTechnique framework that exploits query-level insights from
history to assist tuners in adapting. \dbTechnique organizes historical artifacts into
structured insights, obtains query-level configurations by prompting an LLM with relevant
experiences, and composes them into a holistic configuration with beam search. We evaluate
\dbTechnique's ability to assist state-of-the-art cost-/ML-/LLM-based tuners in adapting to new
environments for OLAP workloads on \pg. Compared to the alternative of continuing to tune from
historical configurations, \dbTechnique assists tuners in finding configurations that improve DBMS
performance by up to 74\% in up to 4.7$\times$ less time.


\newpage
\balance

\bibliographystyle{ACM-Reference-Format}
\bibliography{booster}

\end{document}